\documentclass[twocolumn]{aastex62}

\usepackage{natbib}
\usepackage[style=american]{csquotes}
\newcommand{\sname}{HD\,15337}

\newcommand\gaia{\emph{{\it Gaia}}}

\newcommand\tess{\emph{\it TESS}}

\newcommand\tycho{\emph{\it Tycho}}

\newcommand{\ms}{$\mathrm{m\,s^{-1}}$}
\newcommand{\kms}{$\mathrm{km\,s^{-1}}$}

\newcommand\vsini{$v$\,sin\,$i_\star$}

\newcommand\vmic{$v_{\rm mic}$}
\newcommand\vmac{$v_{\rm mac}$}

\newcommand\teff{$T_{\rm eff}$}

\newcommand\logg{log\,{\it g$_\star$}}

\newcommand{\smass}[1][$M_{\odot}$]{$0.90\pm0.03$ #1}
\newcommand{\sradius}[1][$R_{\odot}$]{$0.856\pm0.017$ #1}
\newcommand{\stemp}[1][$\mathrm{K}$]{ $ 5125\pm50$ #1 }

\newcommand{\Tzerob}[1][days]   {$8411.46156_{-0.00119}^{+0.00094}$~#1} 
\newcommand{\Pb}[1][days]   {$4.75615 \pm 0.00017 $~#1} 
\newcommand{\esinb}[1][ ]   {$0.22_{-0.11}^{+0.09}$~#1} 
\newcommand{\ecosb}[1][ ]   {$0.12_{-0.18}^{+0.14}$~#1} 
\newcommand{\bb}[1][ ]   {$0.33_{-0.14}^{+0.09}$~#1} 
\newcommand{\arb}[1][ ]   {$13.11_{-0.15}^{+0.16}$~#1} 
\newcommand{\rrb}[1][ ]   {$0.01761_{-0.00058}^{+0.00055}$~#1} 
\newcommand{\kb}[1][${\rm m\,s^{-1}}$]   {$3.08_{-0.41}^{+0.44}$~#1} 
\newcommand{\mpb}[1][$M_{\oplus}$]   {$7.51_{-1.01}^{+1.09}$~#1} 
\newcommand{\rpb}[1][$R_{\oplus}$]   {$1.64 \pm 0.06 $~#1} 
 
\newcommand{\eb}[1][ ]   {$0.09 \pm 0.05 $~#1} 
\newcommand{\wb}[1][deg]   {$62_{-32}^{+42}$~#1} 
\newcommand{\ib}[1][deg]   {$88.5_{-0.4}^{+0.6}$~#1} 
\newcommand{\ab}[1][AU]   {$0.0522 \pm 0.0012 $~#1} 
\newcommand{\denpb}[1][${\rm g\,cm^{-3}}$]   {$9.30_{-1.58}^{+1.81}$~#1} 
\newcommand{\Teqb}[1][K]   {$1001 \pm 11.5 $~#1} 
\newcommand{\ttotb}[1][hours]   {$2.49 \pm 0.06 $~#1} 

\newcommand{\Tzeroc}[1][days]   {$8414.5501 \pm 0.0015 $~#1} 
\newcommand{\Pc}[1][days]   {$17.1784 \pm 0.0016 $~#1} 
\newcommand{\esinc}[1][ ]   {$-0.08 _{ - 0.15 } ^ { + 0.16 }$~#1} 
\newcommand{\ecosc}[1][ ]   {$0.12 _{ - 0.17 } ^ { + 0.15 }$~#1} 
\newcommand{\bc}[1][ ]   {$0.87_{-0.02}^{+0.01}$~#1} 
\newcommand{\arc}[1][ ]   {$31.87 \pm  0.70 $~#1} 
\newcommand{\rrc}[1][ ]   {$0.0256 \pm 0.0011 $~#1} 
\newcommand{\kc}[1][${\rm m\,s^{-1}}$]   {$2.16_{-0.45}^{+0.48}$~#1} 
\newcommand{\mpc}[1][$M_{\oplus}$]   {$8.11_{-1.69}^{+1.82}$~#1} 
\newcommand{\rpc}[1][$R_{\oplus}$]   {$2.39 \pm 0.12 $~#1} 
\newcommand{\ec}[1][ ]   {$0.05_{-0.04}^{+0.06}$~#1} 
\newcommand{\wc}[1][deg]   {$329_{-64}^{+69}$~#1} 
\newcommand{\ic}[1][deg]   {$88.5_{-0.1}^{+0.2}$~#1} 
\newcommand{\ac}[1][AU]   {$0.1268 \pm 0.0038 $~#1} 
\newcommand{\denpc}[1][${\rm g\,cm^{-3}}$]   {$3.23_{-0.72}^{+0.90}$~#1} 
\newcommand{\Teqc}[1][K]   {$642 \pm 10 $~#1} 
\newcommand{\ttotc}[1][hours]   {$2.25_{-0.11}^{+0.20 }$~#1}

\newcommand{\Pd}[1][days]   {$36.528 \pm 0.022 $~#1}

\newcommand{\qone}[1][]   {$0.37 \pm 0.08 $~#1} 
\newcommand{\qtwo}[1][]   {$0.25 \pm 0.11 $~#1}

\newcommand{\HARPSone}[1][${\rm m\,s^{-1}}$]   {$-3.8174 \pm 0.0027 $~#1} 
\newcommand{\HARPStwo}[1][${\rm m\,s^{-1}}$]   {$-3.7977 \pm 0.0012 $~#1} 
\newcommand{\jHARPSone}[1][${\rm m\,s^{-1}}$]   {$2.19_{-0.30}^{+0.36}$~#1} 
\newcommand{\jHARPStwo}[1][${\rm m\,s^{-1}}$]   {$2.89_{-0.43}^{+0.58}$~#1} 

\newcommand{\ltrend}[1][${\rm m\,s^{-1}\,d^{-1}}$]   {$-0.0057 \times 0.0017 $~#1} 
\newcommand{\qtrend}[1][${\rm m\,s^{-1}\,d^{-2}}$]   {$-13.3 \pm 2.6 \times 10^{-7} $~#1} 

\received{March 13, 2019}
\revised{March 28, 2019}
\accepted{April 2, 2019}

\shorttitle{The transiting multi-planet system \sname}
\shortauthors{Gandolfi et al.}

\begin{document}

\title{The transiting multi-planet system \sname: two nearly equal-mass planets straddling the radius gap}

\correspondingauthor{Davide Gandolfi}
\email{davide.gandolfi@unito.it}

\author{Davide Gandolfi}
\affiliation{Dipartimento di Fisica, Universit\`a degli Studi di Torino, via Pietro Giuria 1, I-10125, Torino, Italy}

\author{Luca Fossati}
\affiliation{Space Research Institute, Austrian Academy of Sciences, Schmiedlstrasse 6, A-8041 Graz, Austria}

\author{John~H.~Livingston}
\affiliation{Department of Astronomy, University of Tokyo, 7-3-1 Hongo, Bunkyo-ku, Tokyo 113-0033, Japan}

\author{Keivan~G.~Stassun}
\affiliation{Vanderbilt University, Department of Physics \& Astronomy, 6301 Stevenson Center Ln., Nashville, TN 37235, US}
\affiliation{Fisk University, Department of Physics, 1000 17th Ave. N., Nashville, TN 37208, US}

\author{Sascha Grziwa}
\affiliation{Rheinisches Institut f\"ur Umweltforschung an der Universit\"at zu K\"oln, Aachener Strasse 209, 50931 K\"oln, Germany}

\author{Oscar Barrag\'an}
\affiliation{Oxford Astrophysics, Department of Physics, University of Oxford, Denys Wilkinson Building, Keble Road, Oxford, OX1 3RH, UK}

\author{Malcolm Fridlund}
\affiliation{Department of Space, Earth and Environment, Chalmers University of Technology, Onsala Space Observatory, 439 92 Onsala, Sweden}
\affiliation{Leiden Observatory, University of Leiden, PO Box 9513, 2300 RA, Leiden, The Netherlands\label{Leiden}}

\author{Daria Kubyshkina}
\affiliation{Space Research Institute, Austrian Academy of Sciences, Schmiedlstrasse 6, A-8041 Graz, Austria}

\author{Carina~M.~Persson}
\affiliation{Department of Space, Earth and Environment, Chalmers University of Technology, Onsala Space Observatory, 439 92 Onsala, Sweden}

\author{Fei Dai}
\affiliation{Department of Physics and Kavli Institute for Astrophysics and Space Research, MIT, Cambridge, MA 02139, USA}
\affiliation{Department of Astrophysical Sciences, Princeton University, 4 Ivy Lane, Princeton, NJ, 08544, USA}

\author{Kristine~W.\,F.~Lam}
\affiliation{Center for Astronomy and Astrophysics, TU Berlin, Hardenbergstr. 36, 10623 Berlin, Germany}

\author{Simon Albrecht}
\affiliation{Stellar Astrophysics Centre, Dep. of Physics and Astronomy, Aarhus University, Ny Munkegade 120, DK-8000 Aarhus C, Denmark}

\author{Natalie Batalha}
\affiliation{Department of Astronomy and Astrophysics, University of California, Santa Cruz, CA 95064, USA}

\author{Paul G. Beck}
\affiliation{Instituto de Astrof\'\i sica de Canarias, C/\,V\'\i a L\'actea s/n, 38205 La Laguna, Spain}
\affiliation{Departamento de Astrof\'isica, Universidad de La Laguna, 38206 La Laguna, Spain}

\author{Anders Bo Justesen}
\affiliation{Stellar Astrophysics Centre, Dep. of Physics and Astronomy, Aarhus University, Ny Munkegade 120, DK-8000 Aarhus C, Denmark}

\author{Juan Cabrera}
\affiliation{Institute of Planetary Research, German Aerospace Center, Rutherfordstrasse 2, 12489 Berlin, Germany}

\author{Scott Cartwright}
\affiliation{Proto-Logic LLC, 1718 Euclid Street NW, Washington, DC 20009, USA}

\author{William D. Cochran}
\affiliation{Department of Astronomy and McDonald Observatory, University of Texas at Austin, 2515 Speedway,~Stop~C1400,~Austin,~TX~78712,~USA}

\author{Szilard Csizmadia}
\affiliation{Institute of Planetary Research, German Aerospace Center, Rutherfordstrasse 2, 12489 Berlin, Germany}

\author{Misty D. Davies}
\affiliation{NASA Ames Research Center, Moffett Field, CA 94035, USA}

\author{Hans J. Deeg}
\affiliation{Instituto de Astrof\'\i sica de Canarias, C/\,V\'\i a L\'actea s/n, 38205 La Laguna, Spain}
\affiliation{Departamento de Astrof\'isica, Universidad de La Laguna, 38206 La Laguna, Spain}

\author{Philipp Eigm\"uller}
\affiliation{Institute of Planetary Research, German Aerospace Center, Rutherfordstrasse 2, 12489 Berlin, Germany}

\author{Michael Endl}
\affiliation{Department of Astronomy and McDonald Observatory, University of Texas at Austin, 2515 Speedway,~Stop~C1400,~Austin,~TX~78712,~USA}

\author{Anders Erikson}
\affiliation{Institute of Planetary Research, German Aerospace Center, Rutherfordstrasse 2, 12489 Berlin, Germany}

\author{Massimiliano Esposito}
\affiliation{Th\"uringer Landessternwarte Tautenburg, Sternwarte 5, D-07778 Tautenberg, Germany}

\author{Rafael A. Garc\'ia}
\affiliation{Laboratoire AIM, CEA/DSM -- CNRS -- Univ. Paris Diderot - IRFU/SAp, Centre de Saclay, 91191 Gif-sur-Yvette, France}

\author{Robert Goeke}
\affiliation{Department of Physics and Kavli Institute for Astrophysics and Space Research, MIT, Cambridge, MA 02139, USA}

\author{Luc\'ia Gonz\'alez-Cuesta}
\affiliation{Instituto de Astrof\'\i sica de Canarias, C/\,V\'\i a L\'actea s/n, 38205 La Laguna, Spain}
\affiliation{Departamento de Astrof\'isica, Universidad de La Laguna, 38206 La Laguna, Spain}

\author{Eike W. Guenther}
\affiliation{Th\"uringer Landessternwarte Tautenburg, Sternwarte 5, D-07778 Tautenberg, Germany}

\author{Artie P. Hatzes}
\affiliation{Th\"uringer Landessternwarte Tautenburg, Sternwarte 5, D-07778 Tautenberg, Germany}

\author{Diego Hidalgo}
\affiliation{Instituto de Astrof\'\i sica de Canarias, C/\,V\'\i a L\'actea s/n, 38205 La Laguna, Spain}

\author{Teruyuki Hirano}
\affiliation{Department of Earth and Planetary Sciences, Tokyo Institute of Technology, 2-12-1 Ookayama, Meguro-ku, Tokyo 152-8551, Japan}

\author{Maria Hjorth}
\affiliation{Stellar Astrophysics Centre, Dep. of Physics and Astronomy, Aarhus University, Ny Munkegade 120, DK-8000 Aarhus C, Denmark}

\author{Petr Kabath}
\affiliation{Astronomical Institute, Czech Academy of Sciences, Fri\v{c}ova 298, 25165, Ond\v{r}ejov, Czech Republic}

\author{Emil Knudstrup}
\affiliation{Stellar Astrophysics Centre, Dep. of Physics and Astronomy, Aarhus University, Ny Munkegade 120, DK-8000 Aarhus C, Denmark}

\author{Judith Korth}
\affiliation{Rheinisches Institut f\"ur Umweltforschung an der Universit\"at zu K\"oln, Aachener Strasse 209, 50931 K\"oln, Germany}

\author{Jie Li}
\affiliation{Aerospace Computing Inc./NASA Ames Research Center, Moffett Field, CA 94035, USA}

\author{Rafael Luque}
\affiliation{Instituto de Astrof\'\i sica de Canarias, C/\,V\'\i a L\'actea s/n, 38205 La Laguna, Spain}
\affiliation{Departamento de Astrof\'isica, Universidad de La Laguna, 38206 La Laguna, Spain}

\author{Savita Mathur}
\affiliation{Instituto de Astrof\'\i sica de Canarias, C/\,V\'\i a L\'actea s/n, 38205 La Laguna, Spain}
\affiliation{Departamento de Astrof\'isica, Universidad de La Laguna, 38206 La Laguna, Spain}

\author{Pilar Monta\~nes Rodr\'iguez}
\affiliation{Instituto de Astrof\'\i sica de Canarias, C/\,V\'\i a L\'actea s/n, 38205 La Laguna, Spain}
\affiliation{Departamento de Astrof\'isica, Universidad de La Laguna, 38206 La Laguna, Spain}

\author{Norio Narita}
\affiliation{Department of Astronomy, University of Tokyo, 7-3-1 Hongo, Bunkyo-ku, Tokyo 113-0033, Japan}
\affiliation{Instituto de Astrof\'\i sica de Canarias, C/\,V\'\i a L\'actea s/n, 38205 La Laguna, Spain}
\affiliation{Astrobiology Center, NINS, 2-21-1 Osawa, Mitaka, Tokyo 181-8588, Japan}
\affiliation{National Astronomical Observatory of Japan, NINS, 2-21-1 Osawa, Mitaka, Tokyo 181-8588, Japan}
\affiliation{JST, PRESTO, 7-3-1 Hongo, Bunkyo-ku, Tokyo 113-0033, Japan}

\author{David Nespral}
\affiliation{Instituto de Astrof\'\i sica de Canarias, C/\,V\'\i a L\'actea s/n, 38205 La Laguna, Spain}
\affiliation{Departamento de Astrof\'isica, Universidad de La Laguna, 38206 La Laguna, Spain}

\author{Prajwal Niraula}
\affiliation{Department of Earth, Atmospheric and Planetary Sciences, Massachusetts Institute of Technology, Cambridge, MA 02139}

\author{Grzegorz Nowak}
\affiliation{Instituto de Astrof\'\i sica de Canarias, C/\,V\'\i a L\'actea s/n, 38205 La Laguna, Spain}
\affiliation{Departamento de Astrof\'isica, Universidad de La Laguna, 38206 La Laguna, Spain}

\author{Enric Palle}
\affiliation{Instituto de Astrof\'\i sica de Canarias, C/\,V\'\i a L\'actea s/n, 38205 La Laguna, Spain}
\affiliation{Departamento de Astrof\'isica, Universidad de La Laguna, 38206 La Laguna, Spain}

\author{Martin P\"atzold}
\affiliation{Rheinisches Institut f\"ur Umweltforschung an der Universit\"at zu K\"oln, Aachener Strasse 209, 50931 K\"oln, Germany}

\author{Jorge Prieto-Arranz}
\affiliation{Instituto de Astrof\'\i sica de Canarias, C/\,V\'\i a L\'actea s/n, 38205 La Laguna, Spain}
\affiliation{Departamento de Astrof\'isica, Universidad de La Laguna, 38206 La Laguna, Spain}

\author{Heike Rauer}
\affiliation{Center for Astronomy and Astrophysics, TU Berlin, Hardenbergstr. 36, 10623 Berlin, Germany}
\affiliation{Institute of Planetary Research, German Aerospace Center, Rutherfordstrasse 2, 12489 Berlin, Germany}
\affiliation{Institute of Geological Sciences, FU Berlin, Malteserstr. 74-100, D-12249 Berlin}

\author{Seth Redfield}
\affiliation{Astronomy Department and Van Vleck Observatory, Wesleyan University, Middletown, CT 06459, USA}

\author{Ignasi Ribas}
\affiliation{Institut de Ci\`encies de l'Espai (ICE, CSIC), Campus UAB, C/ de Can Magrans s/n, E-08193 Bellaterra, Spain}
\affiliation{Institut d'Estudis Espacials de Catalunya (IEEC), C/ Gran Capit\`a 2-4, E-08034 Barcelona, Spain}

\author{Marek Skarka}
\affiliation{Astronomical Institute, Czech Academy of Sciences, Fri\v{c}ova 298, 25165, Ond\v{r}ejov, Czech Republic}
\affiliation{Department of Theoretical Physics and Astrophysics, Masaryk University, Kotl\'{a}\v{r}sk\'{a} 2, 61137 Brno, Czech Republic}

\author{Alexis M. S. Smith}
\affiliation{Institute of Planetary Research, German Aerospace Center, Rutherfordstrasse 2, 12489 Berlin, Germany}

\author{Pamela Rowden}
\affiliation{School of Physical Sciences, The Open University, Walton Hall, Milton Keynes MK7 6AA, United Kingdom}

\author{Guillermo Torres}
\affiliation{Center for Astrophysics $\vert$ Harvard \& Smithsonian, 60 Garden Street, Cambridge, MA 02138, USA}

\author{Vincent Van Eylen}
\affiliation{Department of Astrophysical Sciences, Princeton University, 4 Ivy Lane, Princeton, NJ, 08544, USA}

\author{Michael L. Vezie}
\affiliation{Department of Physics and Kavli Institute for Astrophysics and Space Research, MIT, Cambridge, MA 02139, USA}

\begin{abstract}
We report the discovery of a super-Earth and a sub-Neptune transiting the star \sname\ (TOI-402, TIC\,120896927), a bright ($V$\,=\,9) K1 dwarf observed by the \emph{Transiting Exoplanet Survey Satellite} (\tess) in Sectors 3 and 4. We combine the \tess\ photometry with archival HARPS spectra to confirm the planetary nature of the transit signals and derive the masses of the two transiting planets. With an orbital period of 4.8 days, a mass of \mpb, and a radius of \rpb, \sname\,b joins the growing group of short-period super-Earths known to have a rocky terrestrial composition. The sub-Neptune \sname\,c has an orbital period of 17.2 days, a mass of \mpc, and a radius of \rpc, suggesting that the planet might be surrounded by a thick atmospheric envelope. The two planets have similar masses and lie on opposite sides of the radius gap, and are thus an excellent testbed for planet formation and evolution theories. Assuming that \sname\,c hosts a hydrogen-dominated envelope, we employ a recently developed planet atmospheric evolution algorithm in a Bayesian framework to estimate the history of the high-energy (extreme ultraviolet and X-ray) emission of the host star. We find that at an age of 150\,Myr, the star possessed on average between 3.7 and 127 times the high-energy luminosity of the current Sun.
\end{abstract}

\keywords{Planetary systems -- Planets and satellites: individual: \sname\,b and c -- Stars: fundamental parameters -- Stars: individual: \sname\ -- Techniques: photometric -- Techniques: radial velocities}

\section{Introduction}

Successfully launched in April 2018, NASA's \emph{Transiting Exoplanet Survey Satellite} (\tess) is making a significant step forward in understanding the diversity of exoplanets, and especially of super-Earths ($R_\mathrm{p}$\,=\,1--2\,R$_\oplus$) and sub-Neptunes ($R_\mathrm{p}$\,=\,2--4\,R$_\oplus$). \tess\ is performing an all-sky photometric search for planets transiting bright stars ($6<V<11$), so that detailed characterizations of the planets and their atmospheres can be performed \citep{Ricker2015}. The survey is broken up into 26 sectors -- each sector being observed for $\sim$28 days and consisting of four cameras with a combined field of view of $24^\circ$$\times$$96^\circ$. Candidate alerts and full-frame images are released every month. As of March 2019, TESS has already announced the discovery of about a dozen transiting planets \citep[see, e.g.,][]{Esposito2019, Gandolfi2018, Huang2018b, Jones2018, Nielsen2019, Quinn2019, Trifonov2019}.

\tess\ has already led to the detection of ``golden'' systems amenable to in-depth characterization of planetary atmospheres, such as $\pi$\,Men, which is a bright ($V$\,=\,5.65) star hosting a transiting super-Earth with a bulk density that is consistent with either a primary, hydrogen-dominated atmosphere, or a secondary, probably CO$_2$/H$_2$O-dominated, atmosphere \citep{Gandolfi2018,Huang2018b}. The discovery of such systems is central for studying planetary atmospheres via multi-wavelength transmission spectroscopy, and for constraining the evolution models of planetary atmospheres.

\tess\ also enables the discovery of multi-planet systems for which both mass and radius can be precisely measured. Because such planets orbit the same star, differences in mean density and atmospheric structure among planets belonging to the same system can be ascribed mainly to differences in planetary mass and orbital separation \citep[see, e.g.,][]{Guenther2017,Prieto-Arranz2018}. This greatly simplifies modeling of their past evolution history, thus constraining how these planets formed \citep{Alibert2005,Alibert2017}. In this respect, even more significant are multi-planet systems in which two or more planets have similar masses, as differences in radii would most likely be due to the different orbital separations.

In this Letter we report the discovery of two small planets transiting the bright ($V$\,=\,9) star \sname\ (Table~\ref{Table:1}), a K1 dwarf observed by \tess\ in Sectors 3 and 4. We combined the \tess\ photometry with archival HARPS radial velocities (RVs) to confirm the planetary nature of the transit signals and derive the masses of the two planets. This Letter is organized as follows. In Sect.~\ref{Sect:Phot}, we describe the \tess\ photometry and the detection of the transit signals. In Sect.~\ref{Sect:HARPS-Observations}, we present the archival HARPS spectra. The properties of the host star are reported in Sect.~\ref{Sect.:StellarFundamentalParameters}. We present the frequency analysis of the HARPS RVs in Sect.~\ref{Sect.:Frequency_Analysis} and the data modeling in Sect.~\ref{Sect.:Joint_Analysis}. Results, discussions, and a summary are given in Sect.~\ref{Sect.:DiscussionConclusions}.

\begin{table}
\footnotesize
\caption{Main identifiers, coordinates, proper motion, parallax, distance, and optical and infrared magnitudes of \sname.}
\label{Table:1}
\begin{tabular}{lrr}
\hline\hline
\noalign{\smallskip}
Parameter & Value & Source \\
\noalign{\smallskip}
\hline
\noalign{\smallskip}
\multicolumn{3}{l}{\emph{Main Identifiers}} \\
\noalign{\smallskip}
HD & 15337 &  \\
HIP & 11433 & \emph{Hipparcos} \\
TIC & 120896927 & TIC$^a$  \\
TOI & 402       & \tess\ Alerts \\
\textit{Gaia} DR2 & 5068777809824976256 & \textit{Gaia} DR2$^b$ \\
\noalign{\smallskip}
\hline
\noalign{\smallskip}
\multicolumn{3}{l}{\emph{Equatorial Coordinates}} \\
\noalign{\smallskip} 
R.A. (J2000.0) & $02^\mathrm{h}\,27^{\mathrm{m}}\,28.3781^{\mathrm{s}}$ & \textit{Gaia} DR2$^b$ \\
Decl. (J2000.0) & $-$27$^{\circ}$\,38$^\prime$\,06.7417${\arcsec}$ & \textit{Gaia} DR2$^b$ \\
\noalign{\smallskip}
\hline
\noalign{\smallskip}
\multicolumn{3}{l}{\emph{Proper Motion, Parallax, and Distance}} \\
$\mu_{\alpha} \cos \delta$ (mas \ yr$^{-1}$) & $-73.590 \pm 0.057$ & \gaia\ DR2$^b$ \\
$\mu_{\delta} $ (mas \ yr$^{-1}$) & $-211.614 \pm 0.082$ & \gaia\ DR2$^b$ \\
Parallax (mas) & $ 22.285\pm0.035$ & \gaia\ DR2$^b$\\ 
Distance (pc) & $44.874\pm0.070$& \gaia\ DR2$^b$ \\ 
\noalign{\smallskip}
\hline
\noalign{\smallskip}
\multicolumn{3}{l}{\emph{Magnitudes}} \\
$B_\mathrm{T}$ & $10.170  \pm 0.027$ & \tycho-2$^c$  \\
$V_\mathrm{T}$ & $9.184   \pm 0.018$ & \tycho-2$^c$  \\
$B$ &  $ 10.009 \pm 0.090$ & APASS$^d$ \\
$V$ &  $ 9.096 \pm 0.017$  & APASS$^d$ \\
$g$ &  $ 9.852 \pm 0.493$  & APASS$^d$ \\
$r$ &  $ 8.847 \pm 0.016$  & APASS$^d$ \\
$i$ &  $ 8.655 \pm 0.054$  & APASS$^d$ \\
$u$ &  $ 11.756\pm 0.075$  & Str\"{o}mgren$^e$ \\
$v$ &  $10.526 \pm 0.046$  & Str\"{o}mgren$^e$ \\
$b$ &  $9.598  \pm 0.032$  & Str\"{o}mgren$^e$ \\
$y$ &  $9.088  \pm 0.030$  & Str\"{o}mgren$^e$ \\
$G$            & $8.8560  \pm 0.0002 $ & \gaia\ DR2$^b$  \\
$G_\mathrm{BP}$& $9.3194 \pm 0.0011 $ & \gaia\ DR2$^b$  \\
$G_\mathrm{RP}$& $ 8.2708 \pm 0.0016 $ & \gaia\ DR2$^b$ \\ 
$J$            & $ 7.553  \pm 0.019  $ & 2MASS$^f$  \\
$H$            & $ 7.215  \pm 0.034  $ & 2MASS$^f$  \\
$K_s$          & $ 7.044  \pm 0.018  $ & 2MASS$^f$  \\
$W1$(3.35 $\mu$m)   &  $6.918 \pm 0.054$  & ALLWISE$^g$ \\
$W2$(4.6 $\mu$m)    &  $7.048 \pm 0.020$  & ALLWISE$^g$ \\
$W3$(11.6 $\mu$m)   &  $7.015 \pm 0.017$  & ALLWISE$^g$ \\
$W4$(22.1 $\mu$m)   &  $6.916 \pm 0.072$  & ALLWISE$^g$ \\
\noalign{\smallskip}
\hline
\end{tabular}
\tablecomments{(a) \tess\ Input Catalog \citep[TIC;][]{Stassun2018b}; (b) \gaia\ Data Release 2 \citep[DR2;][]{GaiaDR2}; (c) \tycho-2 catalog \citep{Hog2000}; (d) AAVSO Photometric All-Sky Survey \citep[APASS;][]{Henden2015}; (e) Str\"omgren-Crawford $uvby\beta$ photometry catalog \citep{Paunzen2015}; (f) Two-micron All Sky Survey \citep[2MASS;][]{Cutri2003}; (g) Wide-field Infrared Survey Explorer catalog \citep[WISE;][]{Cutri2013}.}
\end{table}

~\\
\section{\tess\ photometry}
\label{Sect:Phot}

\begin{figure}
\resizebox{\hsize}{!}{\includegraphics[trim={0 0 0 0},clip]{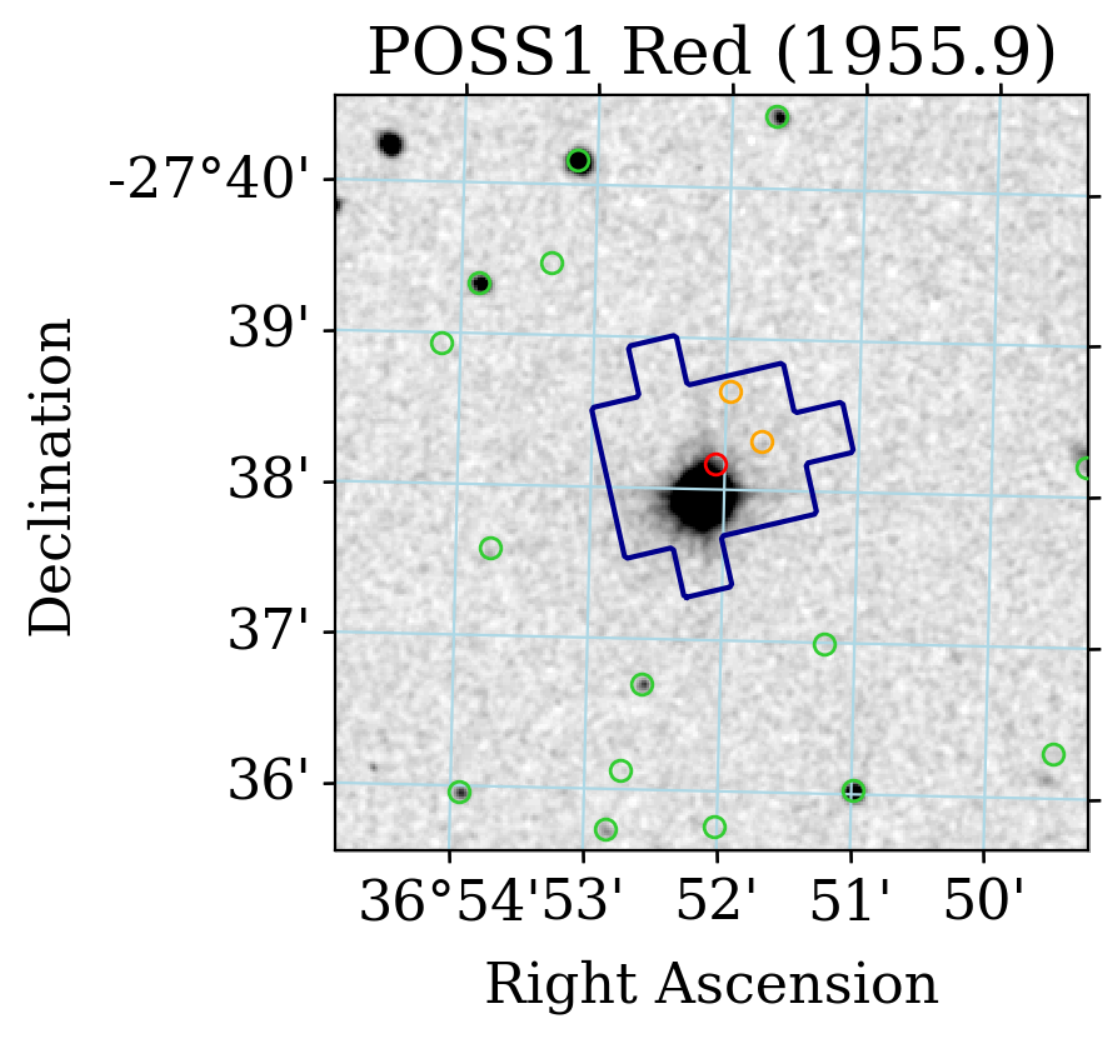}}
\caption{5\,\arcmin\,$\times$\,5\,\arcmin\ 103aE emulsion image taken in 1955 from the First Palomar Sky Survey (POSS1-E), with the Sector 4 SPOC photometric aperture overplotted in blue (\tess\, pixel size is 21\arcsec), and the positions of \gaia\ DR2 sources (J2015.5) within 2\arcmin\ of \sname\ indicated by circles. \sname\ is in red, nearby sources contributing more than 1\% of their flux to the aperture are in orange, and other sources are in green; the Sector 3 aperture is slightly bigger but yields similarly low levels of photometric dilution} (see Sect.~\ref{Sect:Phot}). Due to the proper motion of \sname, there is a $\sim$14\arcsec\ offset between its \gaia\ position and its position in the image.
\label{fig:tessaperture}
\end{figure}

\sname\ (TIC\,120896927) was observed by \tess\ Camera \#2 in Sectors 3 and 4 (charge-coupled devices \#3 and \#4, respectively) from 20 September 2018 to 15 November 2018, and will not be observed further during  the nominal two-year \tess\ mission. Photometry was interrupted when the satellite was re-pointed for data downlink, from BJD$_\mathrm{TDB}=2458395.4$ to BJD$_\mathrm{TDB}=2458396.6$ in Sector 3, and from BJD$_\mathrm{TDB}=2458423.5$ to BJD$_\mathrm{TDB}=2458424.6$ in Sector 4. There is an additional data gap in Sector 4 from BJD$_\mathrm{TDB}=2458418.5$ to BJD$_\mathrm{TDB}=2458421.2$, which was caused by an interruption in communications between the instrument and spacecraft.

\tess\ objects of interest (TOIs) are announced publicly via the \tess\ data alerts web portal\footnote{\url{https://tess.mit.edu/alerts}.} at the Massachusetts Institute of Technology. TOIs 402.01 (\sname\,b) and 402.02 (\sname\,c) were announced on 16 January 2019 and 31 January 2019, respectively, in association with the \sname\ photometry. The \tess\ pixel data and light curves produced by the Science Processing Operations Center \citep[SPOC;][]{Jenkins2016} at NASA Ames Research Center were subsequently made publicly available via the Mikulski Archive for Space Telescopes (MAST).\footnote{\url{https://mast.stsci.edu}.} We iteratively searched the SPOC light curves for transit signals using the Box-least-squares algorithm \citep[BLS;][]{Kovacs2002}, after fitting and removing stellar variability using a cubic spline with knots every 1.0 day. We recovered two signals corresponding to the TOIs, but no other significant signals were detected. We also tried removing variability using the wavelet-based filter routines {\tt VARLET} and {\tt PHALET}, but it did not change the BLS results; we are thus confident that the two signals are robustly detected and are not the result of data artifacts resulting from the choice of variability model or residual instrumental systematic signals.

The SPOC light curves are produced using automatically selected optimal photometric apertures. We also produced light curves from the \tess\ pixel data using a series of apertures \citep{Gandolfi2018, Esposito2019}, and found that apertures larger than the SPOC aperture shown in Fig.\,\ref{fig:tessaperture} minimized the 6.5 hr combined differential photometric precision (CDPP) noise metric \citep{2012PASP..124.1279C}. However, the transit signals recovered from these light curves were slightly less significant, which we attributed to the improvement in light curve quality afforded by the Presearch Data Conditioning \citep[][]{2012PASP..124.1000S,2012PASP..124..985S} pipeline used by the SPOC, which corrects for common-mode systematic noise; for this reason, we opted to analyze the SPOC~light curves for the remainder of the analysis in this Letter.

To investigate the possibility of diluting flux from stars other than \sname, we visually inspected archival images and compared \gaia\ DR2 \citep{GaiaDR2} source positions with the SPOC photometric apertures. We used the coordinates of \sname\ from the \tess\ Input Catalog\footnote{Available at \url{https://mast.stsci.edu/portal/Mashup/Clients/Mast/Portal.html}.} \citep[][]{Stassun2018b} to retrieve \gaia\ DR2 sources using a search radius of 3\arcmin. In an archival image taken in 1955 from the Firs Palomar Sky Survey (POSSI-E)\footnote{Available at \url{http://archive.stsci.edu/cgi-bin/dss_form}.}, \sname\ is offset from its current position by $\sim$14\arcsec\ due to proper motion, but this is not sufficient to completely rule out chance alignment with a background source; however, such an alignment with a bright source is highly unlikely. Assuming the \tess\ point spread function (PSF) can be approximated by a two-dimensional Gaussian profile with a full-width at half maximum (FWHM) of $\sim$25\arcsec, we found that 98.9\% (98.5\%) of the flux from \sname\ is within the Sector 3 (Sector 4) SPOC aperture. Approximating the \tess\ bandpass with the \gaia\ $G_\mathrm{RP}$ bandpass, the transit signals from \sname\ should be diluted by less than 0.01\% in both apertures; \sname\ is the only star bright enough to be the source of the transit signals. Two other \gaia\ DR2 sources (5068777809825770112 and 5068777745400963584) also contribute flux, but they are too faint to yield significant dilution ($G_\mathrm{RP} \approx 19$ mag). Fig.~\ref{fig:tessaperture} shows the archival image, along with \gaia\, DR2 source positions and the Sector 4 SPOC photometric aperture.

\section{HARPS spectroscopic observations}
\label{Sect:HARPS-Observations}

\sname\ was observed between 15 December 2003 and 06 September 2017 UT with the High Accuracy Radial velocity Planet Searcher (HARPS) spectrograph \citep[R\,$\approx$\,115\,000,][]{Mayor2003} mounted at the ESO-3.6\,m telescope, as part of the observing programs 072.C-0488, 183.C-0972, 192.C-0852, 196.C-1006, and 198.C-0836. We retrieved the publicly available reduced spectra from the ESO archive, along with the cross-correlation function (CCF) and its bisector, computed from the dedicated HARPS pipeline using a K5 numerical mask \citep{Baranne1996}. On June 2015, the HARPS fiber bundle was upgraded and a new set of octagonal fibers, with improved mode-scrambling capabilities, were installed \citep{LoCurto2015}. To account for the RV offset caused by the instrument refurbishment, we treated the HARPS RVs taken before/after June 2015 as two different data sets. Tables~\ref{Table:HARPS1} and \ref{Table:HARPS2} list the HARPS RVs taken with the old and new fiber bundle, along with the RV uncertainties, the full-width at half maximum (FWHM) and bisector span (BIS) of the CCF, the exposure times, and the signal-to-noise ratio (S/N) per pixel at 5500\,\AA. Time stamps are given in barycentric Julian Date in the barycentric dynamical time (BJD\,$_\mathrm{TDB}$). We rejected two data points -- marked with asterisks in Tables~\ref{Table:HARPS1} and \ref{Table:HARPS2} -- because of poor S/N ratio (BJD$_\mathrm{TDB}$\,=\,2455246.519846) or systematics (BJD$_\mathrm{TDB}$\,=\,2457641.794439).

\section{Stellar fundamental parameters}
\label{Sect.:StellarFundamentalParameters}

\subsection{Spectroscopic parameters}
\label{Sect:SpecParam}

We co-added the HARPS spectra obtained with the old and new fiber bundle separately to get two combined spectra with S/N per pixel at 5500\,\AA\ of 590 (old fiber) and 490 (new fiber). We derived the spectroscopic parameters of \sname\ from the co-added HARPS spectra using Spectroscopy Made Easy ({\tt SME}), a spectral analysis tool that calculates synthetic spectra and fits them to high-resolution observed spectra using a $\chi^2$ minimizing procedure. The analysis was performed with the non-local thermodynamic equilibrium (non-LTE) \texttt{SME} version 5.2.2, along with \texttt{MARCS} model atmospheres \citep{Gustafsson2008}. 

We estimated a microturbulent velocity of \vmic=$0.80\pm0.10$~\kms\ from the empirical calibration equations for Sun-like stars from \cite{Bruntt2010b}. The effective temperature \teff\ was measured fitting the wings of the H$_\alpha$ and H$_\beta$ lines, as well as the Na\,{\sc i} doublet at 5890 and 5896~\AA\ \citep{Fuhrmann93,Axer94,Fuhrmann94,Fuhrmann97a,Fuhrmann97b}. The surface gravity \logg\ was determined from the wings of the Ca\,{\sc i}~$\lambda$\,6102, $\lambda$\,6122, $\lambda$\,6162\,\AA\ triplet, and the Ca\,{\sc i} $\lambda$\,6439\,\AA\ line, as well as from the Mg\,{\sc i}~$\lambda$\,5167, $\lambda$\,5173, $\lambda$\,5184\,\AA\ triplet. We measured the iron abundance [Fe/H], the macroturbulent velocity \vmac, and the projected rotational velocity \vsini\ by simultaneously fitting the unblended iron lines in the spectral region 5880--6600\,\AA.

Our analyses applied to the two stacked HARPS spectra provided consistent results, well within the uncertainties. The final adopted values are listed in Table~\ref{tab:parameters}. We derived an effective temperature of \teff=5125$\pm$50~K, surface gravity \logg=$4.40\pm0.10$\,(cgs), and an iron abundance relative to solar of [Fe/H]=$0.15\pm0.08$~dex. We also measured a calcium abundance of [Ca/H]=$0.16\pm0.05$~dex and a sodium abundance of [Na/H]=$0.27\pm0.09$~dex. We found a macroturbulent velocity of \vmac=$3.0\pm1.0$\,km/s in agreement with the value predicted from the empirical equations of \citet{Doyle2014}. The projected rotational velocity was found to be \vsini=$1.0\pm1.0$~\kms.

\subsection{Stellar mass, radius, age and interstellar extinction}
\label{Sect:Mass_Radius_Age_Av}

\begin{figure}
\resizebox{\hsize}{!}{\includegraphics[trim={5.0cm 2cm 3.0cm 2cm},clip]{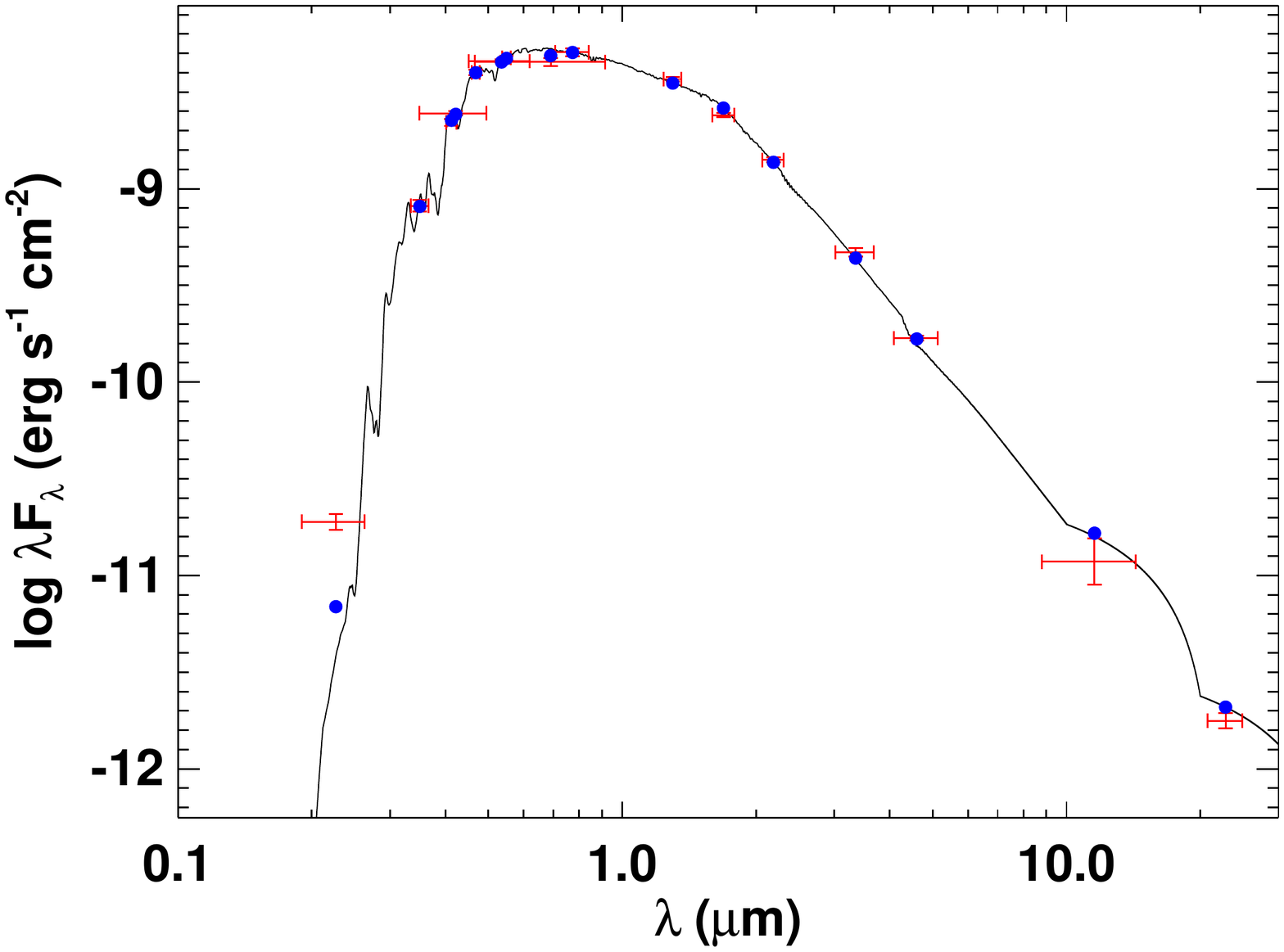}}
\caption{Spectral energy distribution (SED) of \sname. Red symbols represent the observed photometric measurements, where the horizontal bars represent the effective width of the passband. Blue symbols are the model fluxes from the best-fit Kurucz atmosphere model (black). 
\label{Fig:SED}}
\end{figure}

\begin{figure}
\begin{center}
\resizebox{\hsize}{!}{\includegraphics[trim={2.0cm 2.0cm 1.5cm 4.6cm},clip,width=\columnwidth]{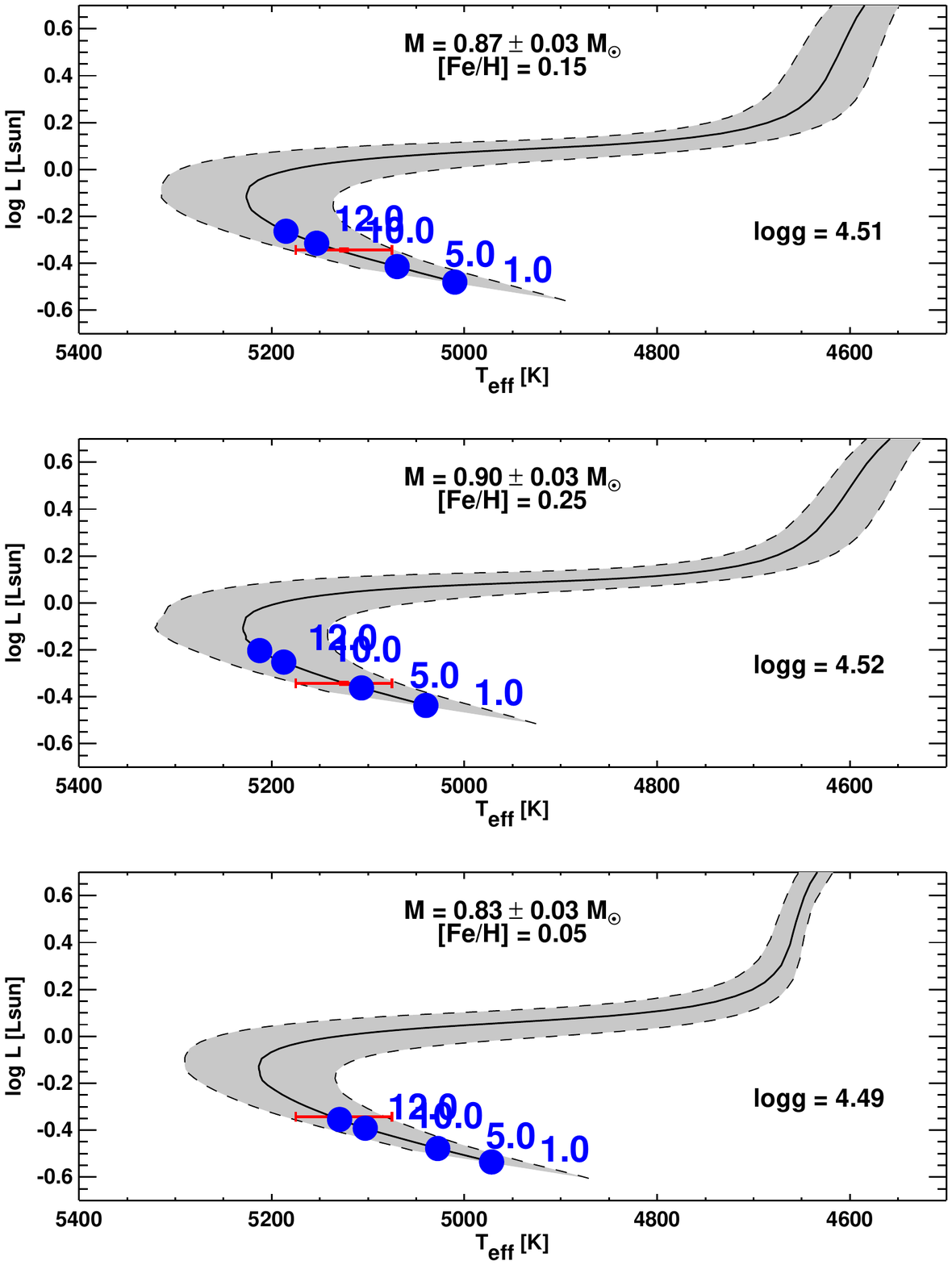}}
\caption{Hertzsprung--Russell Diagram for HD~15337 based on the observed effective temperature and bolometric luminosity, the latter computed directly from $F_{\rm bol}$ and the {\it Gaia\/} parallax-based distance. Each panel compares the observed properties of the star to evolutionary tracks from the Yonsei--Yale models  \citep{Yi2001,Spada2013} for different permitted combinations of stellar mass and metallicity. Blue points with labels represent the model ages in Gyr. The central panel represents the case most compatible with all of the available data, including the stellar age of $\approx$5.1~Gyr as determined from the observed chromospheric activity and stellar rotation period (see the text). 
\label{Fig:HR}}
\end{center}
\end{figure}

We performed an analysis of the broadband spectral energy distribution (SED) together with the \gaia\ Data Release 2 \citep[DR2;][]{GaiaDR2} parallax in order to determine an empirical measurement of the stellar radius, following the procedures described in \citet{Stassun2016}, \citet{Stassun2017}, and \citet{Stassun2018a}. We retrieved the $B_\mathrm{T}$ and $V_\mathrm{T}$ magnitudes from \tycho-2 catalog \citep{Hog2000}, the Str\"{o}mgren $ubvy$ magnitudes from \citet{Paunzen2015}, the $BVgri$ magnitudes from APASS \citep{Henden2015}, the $JHK_S$ magnitudes from {\it 2MASS} \citep{Cutri2003}, the $W1$--$W4$ magnitudes from {\it ALLWISE} \citep{Cutri2013}, and the $G$ magnitude from \gaia\ DR2 \citep{GaiaDR2}. Together, the available photometry spans the full stellar SED over the wavelength range 0.35--22~$\mu$m (Table~\ref{Table:1} and Fig.~\ref{Fig:SED}). In addition, we retrieved the near-ultraviolet (NUV) flux from the \emph{Galaxy Evolution Explorer} ({\it GALEX}) survey \citep{Bianchi2011} in order to assess the level of chromospheric activity, if any. 

We performed a fit using Kurucz stellar atmosphere models \citep{Castelli2003}, with the fitted parameters being the effective temperature $T_{\rm eff}$ and iron abundance [Fe/H], as well as the interstellar extinction $A_\mathrm{v}$, which we restricted to the maximum line-of-sight value from the dust maps of \citet{Schlegel1998}. The broadband SED is largely insensitive to the surface gravity (\logg), thus we simply adopted the value from the initial spectroscopic analysis presented in the previous subsection. The resulting fit (Fig.~\ref{Fig:SED}) gives a reduced $\chi^2$ of 2.3 (excluding the {\it GALEX} NUV flux, which is consistent with a modest level of chromospheric activity). The best-fitting effective temperature and iron content are \teff\,=\,5130\,$\pm$\,50~K and [Fe/H]\,=\,$0.1_{-0.1}^{+0.2}$\,dex, respectively, in excellent agreement with the spectroscopic values (Sect.~\ref{Sect:SpecParam} and Table~\ref{tab:parameters}). We found that the reddening of \sname\ is consistent with zero ($A_\mathrm{v}$\,=\,0.02\,$\pm$\,0.02~mag), as expected given the relatively short distance to the star ($\sim$45\,pc). Integrating the unreddened model SED gives a bolometric flux at Earth of $F_{\rm bol} =  7.29 \pm 0.08 \times 10^{-9}$ erg~s~cm$^{-2}$. Taking the $F_{\rm bol}$ and \teff\ together with the \gaia\ DR2 parallax, adjusted by $+0.08$~mas to account for the systematic offset reported by \citet{StassunTorres2018}, gives the stellar radius as $R_\star$\,=\,0.856\,$\pm$\,0.017\,$R_\odot$. Finally, estimating the stellar mass from the empirical relations of \citet{Torres2010} and a 6\% error from the empirical relation itself gives a stellar mass of $M_\star$\,=\,0.91\,$\pm$\,0.06\,$M_\odot$. 

We can refine the stellar mass estimate by taking advantage of the observed chromospheric activity, which can constrain the age of the star via empirical relations. For example, taking the chromospheric activity indicator, $\log R'_\mathrm{HK}$\,=\,$-4.916$\,$\pm$0.038 from \citet{GomesDaSilva2014} and applying the empirical relations of \citet{Mamajek2008}, gives a predicted age of 5.1\,$\pm$\,0.8~Gyr. As shown in Fig.\,\ref{Fig:HR}, according to the Yonsei--Yale stellar evolutionary models \citep{Yi2001,Spada2013}, this age is most compatible with a stellar mass of $M_\star$\,=\,0.90\,$\pm$\,0.03\,$M_\odot$ and [Fe/H]\,=\,0.25 dex, which with the empirically determined stellar radius implies a stellar \logg\,=\,4.53\,$\pm$\,0.02 (cgs) -- in good agreement with the spectroscopic value of \logg\,=\,4.40\,$\pm$\,0.10\,(cgs).

\begin{figure}
\resizebox{\hsize}{!}{\includegraphics[trim={0.0cm 0cm 0.0cm 0cm},clip]{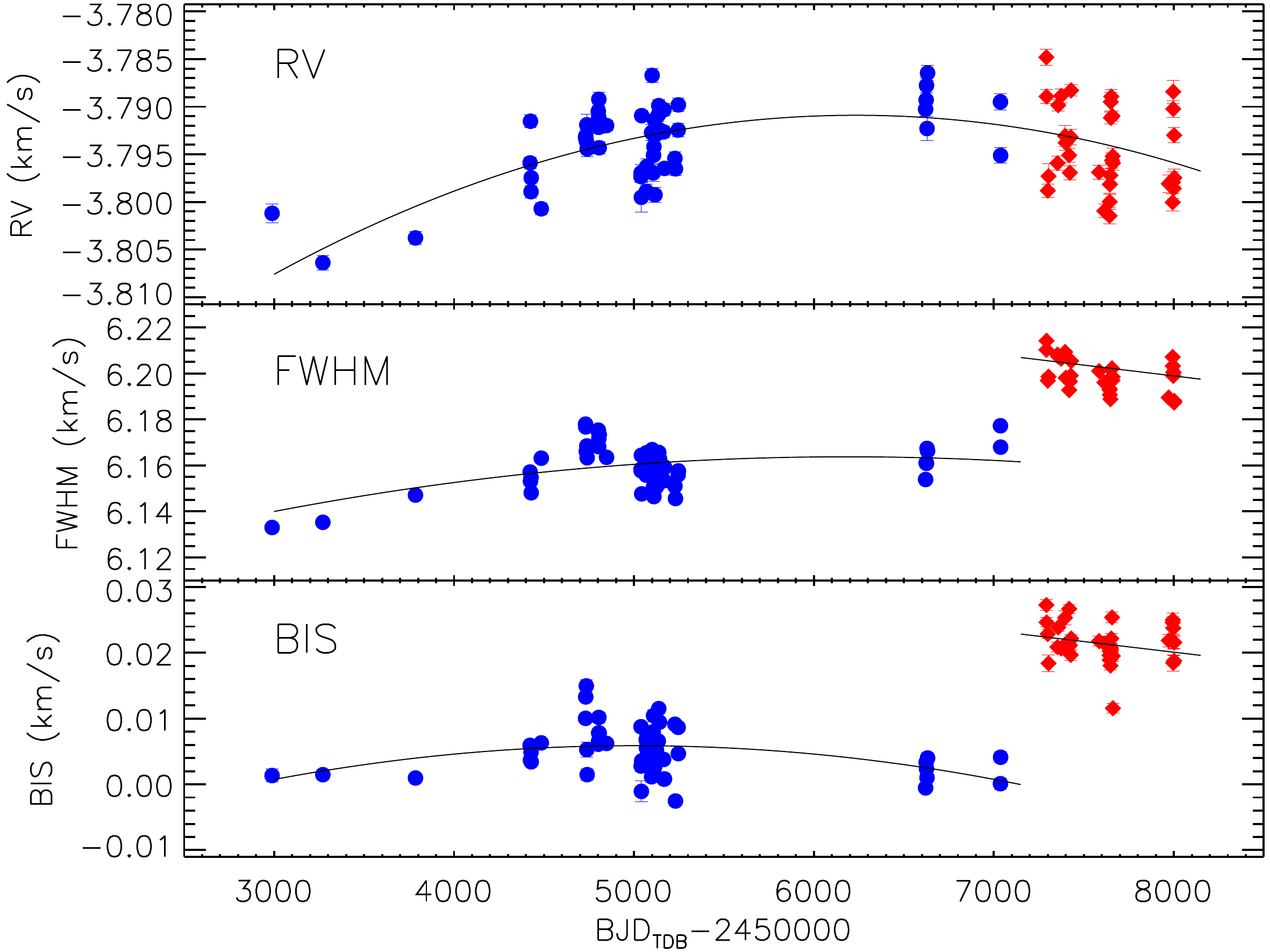}}
\caption{Offset-corrected HARPS RVs of \sname\ (upper panel), and FWHM and BIS of the cross-correlation function (middle and lower panels). The blue circles and red diamonds mark the measurements acquired with the old and new fiber bundle, respectively. The thick lines mark the best-fitting parabolic curves to the data (see the text). \label{Fig:HARPS_RVs}}
\end{figure}

Other combinations of stellar mass and metallicity are compatible with the observed effective temperature and radius (Fig.~\ref{Fig:HR}); however they require ages that are incompatible with that predicted by the chromospheric $R'_\mathrm{HK}$ emission. Finally, we can further corroborate the activity-based age estimate by also using empirical relations to predict the stellar rotation period from the activity. For example, the empirical relation between $R'_\mathrm{HK}$ and rotation period from \citet{Mamajek2008} predicts a rotation period for this star of $\approx$42~days, which is compatible with the observed rotation period derived from the HARPS RVs and activity indicators ($P_\mathrm{rot}$\,=\,36.5 days; see the following section).

\section{Frequency analysis of the HARPS measurements}
\label{Sect.:Frequency_Analysis}

\begin{figure*}[ht]
\resizebox{\hsize}{!}{\includegraphics[trim={0.0cm 0cm 0.0cm 0cm},clip]{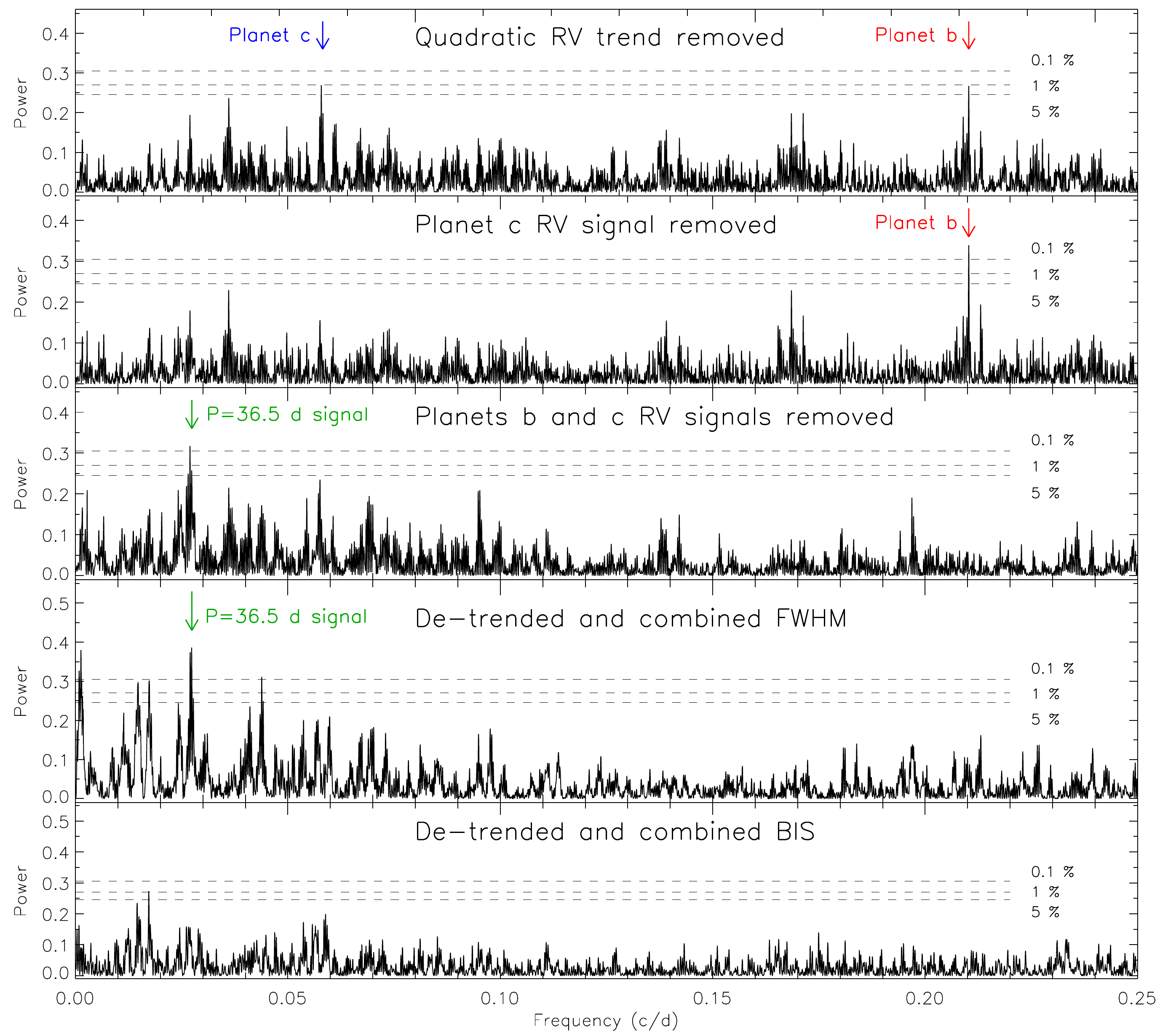}}
\caption{Generalized Lomb--Scargle periodograms of (1) the combined HARPS RV measurements, following the subtraction of the quadratic trend (first panel); (2) the RV residuals after subtracting the signal of \sname\,c (second panel); (3) the RV residuals after subtracting the signal of \sname\,b and c (third panel); (4) the FWHM of the cross-correlation function (fourth panel); (5) the bisector span (BIS) of the cross-correlation function (fifth panel). The dashed horizontal lines mark the FAP at 0.1\%, 1\% and 5~\%. The frequencies of the two transiting planets, as well as of the signal at 36.5 days are marked with vertical arrows. \label{Fig:GLS}}
\end{figure*}

We performed a frequency analysis of the HARPS time-series to search for the Doppler reflex motion induced by the two transiting planets discovered by \tess. We accounted for the RV offset between the two different set-ups of the instrument (old and new fiber bundle) using the value of 19.7\,\ms\ derived from the joint analysis presented in Sect.~\ref{Sect.:Joint_Analysis}, which is in good agreement with the expected offset for a slowly rotating K1\,V star such as \sname\ \citep{LoCurto2015}. 

The offset-corrected HARPS RVs are displayed in Fig.\,\ref{Fig:HARPS_RVs} (upper panel), along with the time-series of the full-width at half maximum (FWHM; middle panel) and bisector span (BIS; lower panel). The generalized Lomb--Scargle (GLS) periodogram \citep{Zechmeister2009} of the combined RV data shows significant power at frequencies lower than the inverse of the temporal baseline of the HARPS observations, which is visible as a quadratic trend in the upper panel of Fig.\,\ref{Fig:HARPS_RVs}. A similar trend is observed in the FWHM obtained with the old fiber bundle (middle panel, blue circles), suggesting that the RV trend might be due to long-term stellar variability (e.g., magnetic cycles)\footnote{We note that the FWHM and BIS offsets between the two instrument set-ups are unknown.}. Alternatively, the RV trend might be induced by a long period orbiting companion, while the long-term variation of the FWHM might be ascribable to the steady instrument de-focusing observed between 2004 and 2015 \citep{LoCurto2015}.

The upper panel of Fig.~\ref{Fig:GLS} shows the GLS periodogram of the combined HARPS RVs, following the subtraction of the best-fitting quadratic trend (cfr. Fig.\,\ref{Fig:HARPS_RVs}). The peaks with the highest power are found at the orbital frequencies of the two transiting planets ($f_c$\,=\,0.058~c/d and $f_b$\,=\,0.210~c/d), with false-alarm probabilities\footnote{The FAP was derived using the bootstrap method described in \citep{Kuerster1997}.} (FAPs) of $\approx$\,1\% and RV semi-amplitude of about 2.0-2.5\,\ms. The periodogram of the RV residuals after subtracting the signal of the outer planet (Fig.~\ref{Fig:GLS}, second panel), shows a significant peak (FAP\,$<$\,0.1\,\%) at the frequency of the inner planet. The two peaks have no counterparts in the periodograms of the activity indicators \footnote{We combined the activity indicators from the two HARPS fibers by subtracting the best-fitting second-order polynomials shown in Fig.~\ref{Fig:HARPS_RVs}.} (FWHM and BIS; Fig.\,\ref{Fig:GLS}, fourth and fifth panels), suggesting that the signals are induced by two orbiting planets with periods of 4.8 and 17.2\,days. Finally, the GLS periodogram of the RV residuals after subtracting the quadratic trend and the signals of the two planets (Fig.\,\ref{Fig:GLS}, third panel) displays a peak with a FAP\,$<$\,0.1\,\% at $\sim$36.5 days, which is also significantly detected in the periodogram of the FWHM (fourth panel). We interpreted the 36.5 days signal as the rotation period of the star, which agrees with the value expected from the $R'_\mathrm{HK}$ activity indicator (Sect.~\ref{Sect:Mass_Radius_Age_Av}).

\section{Joint analysis}
\label{Sect.:Joint_Analysis}

We performed a joint analysis of the \tess\ light curve (Sect.~\ref{Sect:Phot}) and RV measurements (Sect.~\ref{Sect:HARPS-Observations}) using the software suite \texttt{pyaneti}, which allows for parameter estimation from posterior distributions calculated using Markov chain Monte Carlo (MCMC) methods.

We removed stellar variability from the \tess\ light curves using a cubic spline with knots spaced every 1.5 days. We then extracted $\sim$8 hours of \tess\ photometry centered on each of the nine (\sname\,b) and three (\sname\,c) transits observed by \tess\ during Sectors 3 and 4. As described in Sect.~\ref{Sect:HARPS-Observations}, we rejected two HARPS RVs and used the remaining 85 Doppler measurements, while accounting for an RV offset between the two different HARPS set-ups.

The RV model includes a linear and a quadratic term, to account for the long-term variation described in Sect.~\ref{Sect.:Frequency_Analysis}, as well as two Keplerians, to account for the Doppler reflex motion induced by \sname\,b and \sname\,c. The RV stellar signal at the star's rotation period was modeled as an additional coherent sine-like curve whose period was constrained with a uniform prior centered at $P_\mathrm{rot}$\,=\,36.5\,days and having a width of 0.2~day, as derived from the FWHM of the peak detected in the periodogram of the HARPS FWHMs. For the phase and amplitude of the activity signal we adopted uniform priors. While this simple model might not fully reproduce the periodic and quasi-periodic variations induced by evolving active regions carried around by stellar rotation, it has proven to be effective in accounting for the stellar signal of active and moderately active stars \citep[e.g.,][]{Pepe2013,Gandolfi2017,Barragan2018,Prieto-Arranz2018}. Any variation not properly modeled by the coherent sine-curve, and/or any instrumental noise not included in the nominal RV uncertainties, were accounted for by fitting two RV jitter terms for the two HARPS set-ups.

\begin{figure*}[th]
\includegraphics[height=0.29\textwidth]{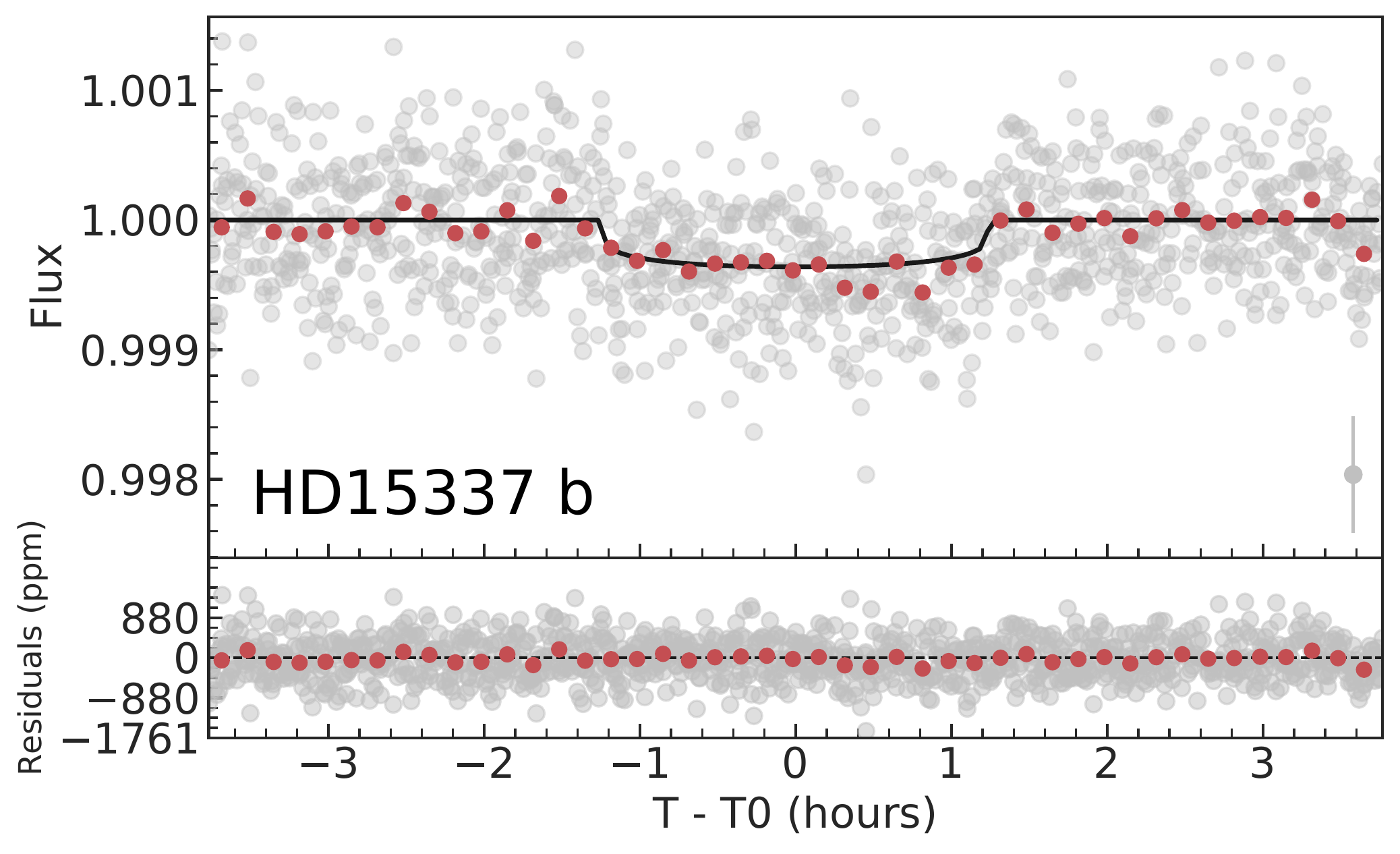} 
\includegraphics[height=0.29\textwidth]{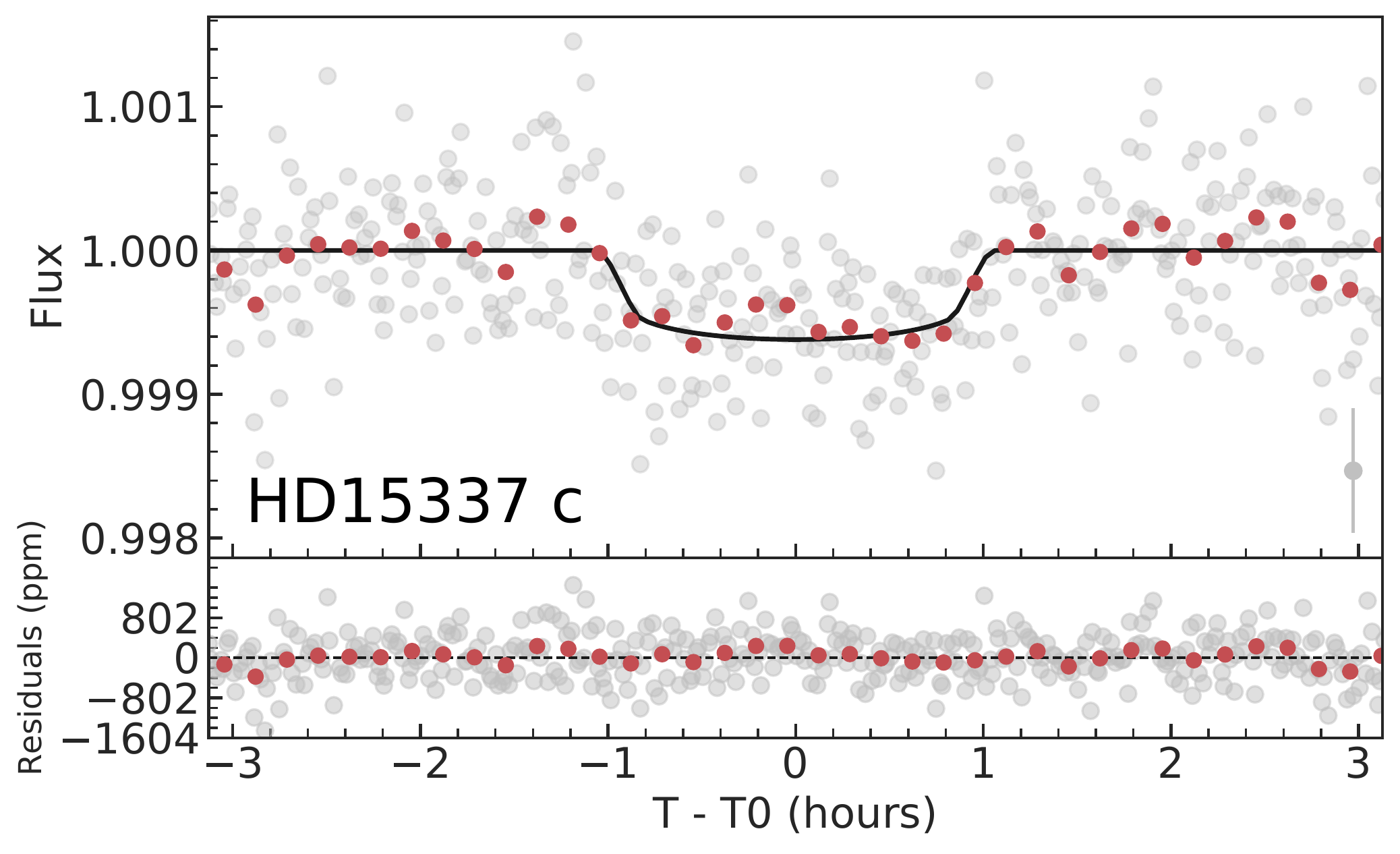} 
\centering
\caption{Folded transit light curves of \sname\,b (left panel) and \sname\,c (middle panel), based on nice and three single transits observed by \tess. The best-fitting transit models are overplotted with thick black lines. The \tess\ data points are shown with gray circles, whereas the 10 minutes binned data are displayed with red circles. 
\label{Fig:TransitLightcurves} }
\end{figure*}  

\begin{figure*}
\includegraphics[height=0.350\textwidth]{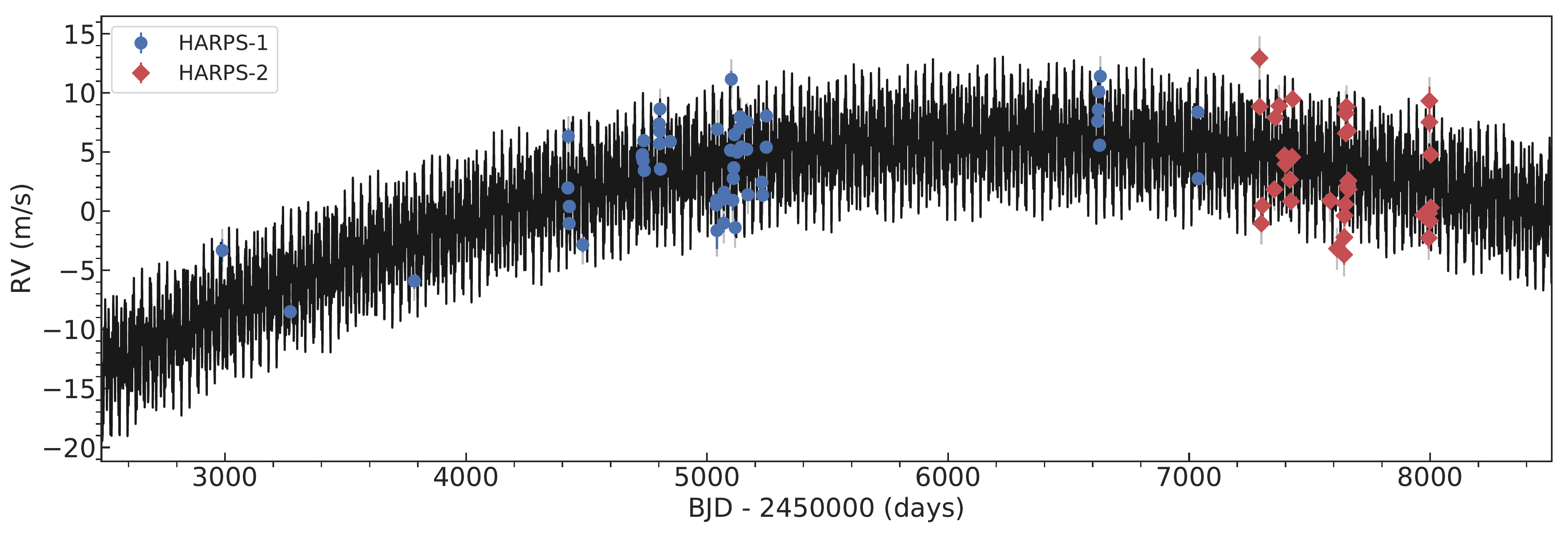} \includegraphics[height=0.2077\textwidth]{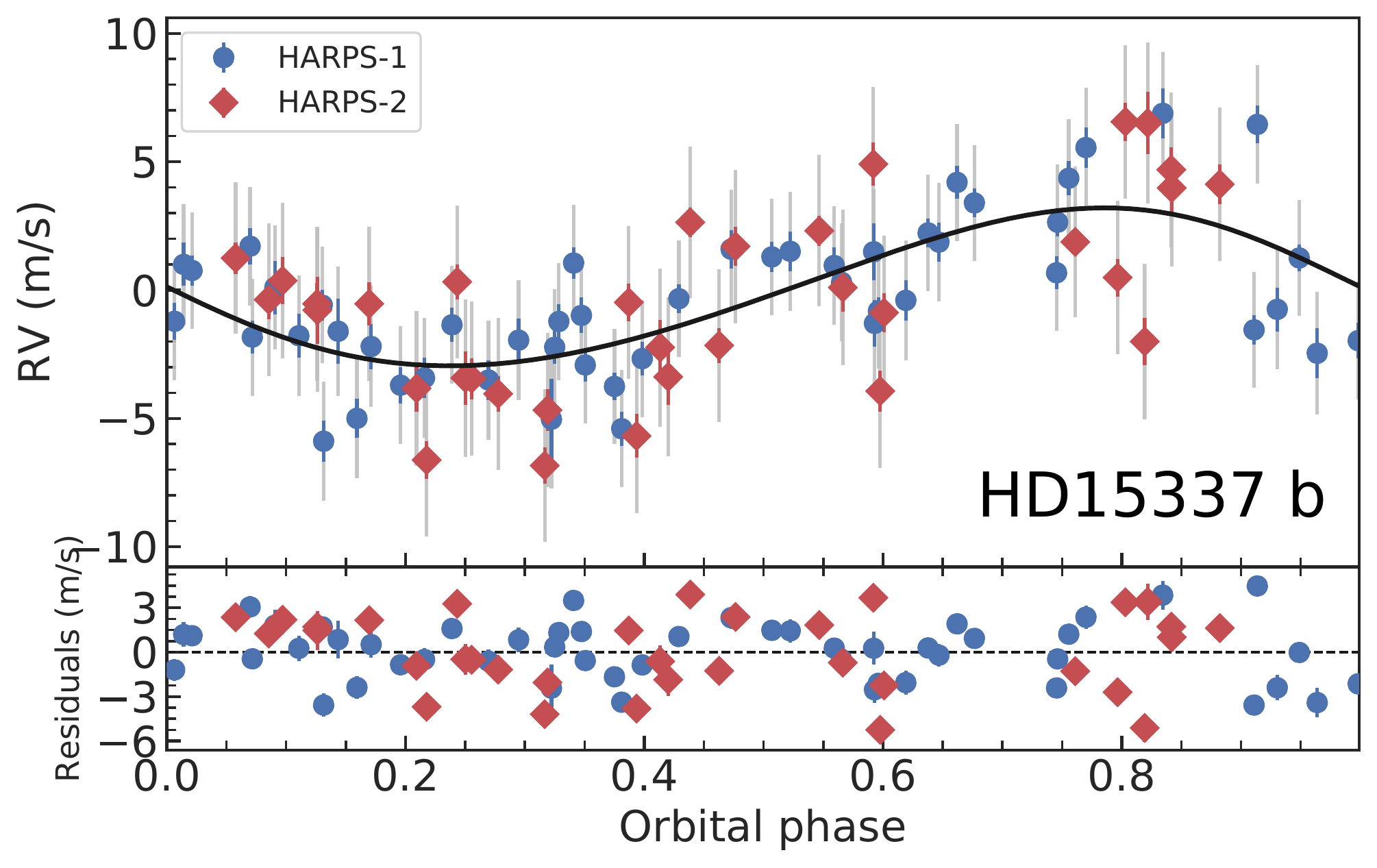} \includegraphics[height=0.2077\textwidth]{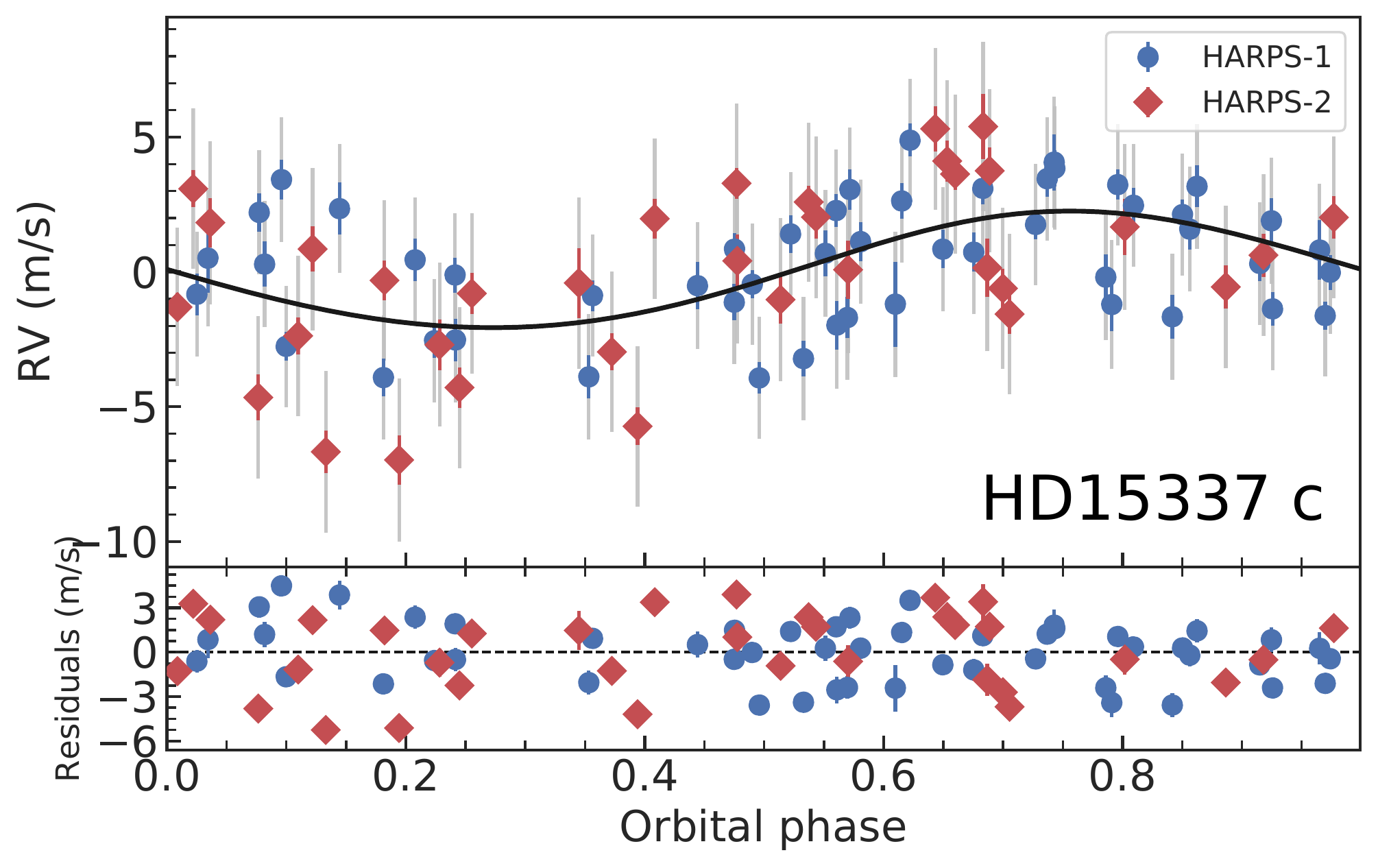}
\includegraphics[height=0.2077\textwidth]{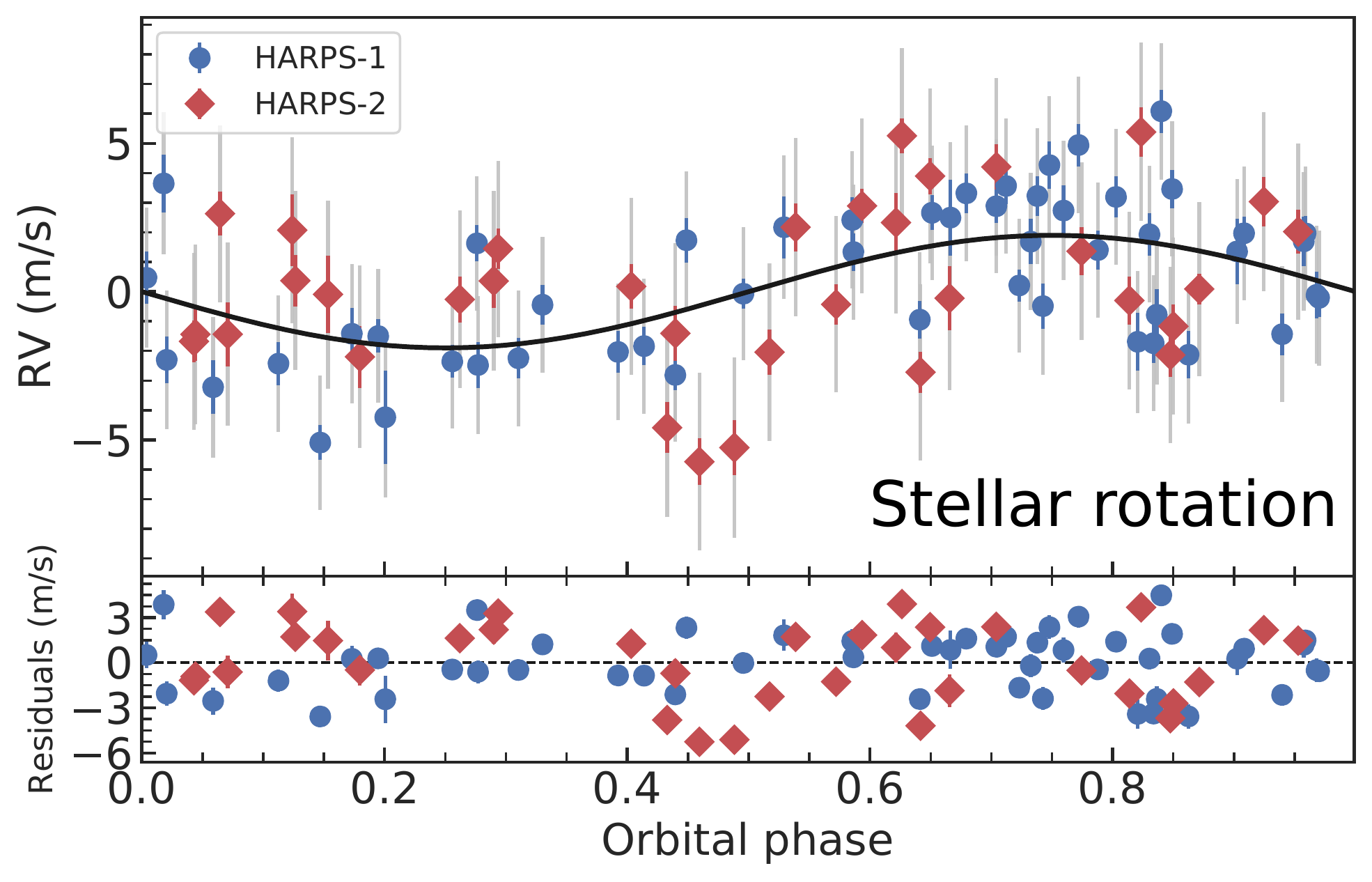}
\caption{Upper panel: HARPS RV measurements vs. time, following the subtraction of the systemic velocities derived for the old (blue circles) and new (red diamonds) instrument set-up. Lower panels: phase-folded RV curves of \sname\,b (left), \sname\,c (middle) and stellar signal at 36.5 days (right). The best-fitting Keplerian and sine models are overplotted with thick black lines. The vertical gray lines mark the error bars including the RV jitter. \label{Fig:RVCurves} }
\end{figure*} 

We modeled the \tess\ transit light curves using the limb-darkened quadratic model of \citet{2002ApJ...580L.171M}. For the limb-darkening coefficients, we set Gaussian priors using the values derived by \citet{Claret2017} for the \tess\ passband. We imposed conservative error bars of 0.1 on both the linear and quadratic limb-darkening terms. For the eccentricity and argument of periastron we adopted the parametrization proposed by \citet{Anderson2011}. A preliminary analysis showed that the transit light curve poorly constrains the scaled semi-major axis ($a/R_\star$). We therefore set a Gaussian prior on $a/R_\star$ using \emph{Kepler}'s third law, the orbital period, and the derived stellar mass and radius (Sect.~\ref{Sect:Mass_Radius_Age_Av}). We imposed uniform priors for the remaining fitted parameters. Details of the fitted parameters and prior ranges are given in Table~\ref{tab:parameters}. We used 500 independent Markov chains initialized randomly inside the prior ranges. Once all chains converged, we used the last 5000 iterations and saved the chain states every 10 iterations. This approach generates a posterior distribution of 250,000 points for each fitted parameter. Table~\ref{tab:parameters} lists the inferred planetary parameters. They are defined as the median and 68\% region of the credible interval of the posterior distributions for each fitted parameter. The transit and RV curves are shown in Fig.~\ref{Fig:TransitLightcurves} and Fig.~\ref{Fig:RVCurves}, along with the best-fitting models.

We also experimented with Gaussian Processes (GPs) to model the correlated RV noise associated with stellar activity. GPs model stochastic processes with a parametric description of the covariance matrix. GP regression has proven to be successful in modeling the effect of stellar activity for several other exoplanetary systems \citep[see, e.g.,][]{Haywood2014,Grunblatt2015,LM2016,Barragan2018}. To this aim, we modified the code \texttt{pyaneti} in order to include a GP algorithm coupled to the MCMC method. We implemented the GP approach proposed by \citet{Rajpaul2015}. Briefly, this framework assumes that the star-induced RV variations and activity indicators can be modeled by the same underlying GP and its derivative. This allows the GP to disentangle the RV activity component from the planetary signals.

We assumed that the stellar activity can be modeled by the quasi-periodic kernel described by \citet{Rajpaul2015}. We modeled together the HARPS RV, BIS, and FWHM time-series and we treated RV and BIS as being described by the GP and its first derivative, while for FWHM we assumed that it is only described by the GP. The fitted hyper-parameters are then $V_c$, $V_r$, $B_c$, $B_r$, $L_c$, as defined by \citet{Rajpaul2015}, to account for the GP amplitudes of the RV, BIS, and FWHM signals, the period of the activity signal $P_{\rm GP}$, the inverse of the harmonic complexity $\lambda_{\rm p}$, and the long term evolution timescale $\lambda_{\rm e}$. We coupled this GP approach with the joint modeling described in the previous paragraphs of the present section (omitting the extra coherent signal).

As for the planetary signals, we imposed the same priors listed in Table~\ref{tab:parameters}. For the hyper-parameters, we used uniform priors, except for $P_{\rm GP}$, for which we imposed a Gaussian prior with mean 36.5\,days and standard deviation of 0.2\,day. We used 250 chains to explore the parameter space. We created the posterior distributions with 500 iterations of converged chains, which generated a posterior distribution with 250,000 points for each parameters.

For planets b and c we derived an RV semi-amplitude of $2.71_{-0.51}^{+0.54}$~\ms and $2.06_{-0.58}^{+0.64}$~\ms, respectively, which are in very good agreement with the values reported in Table~\ref{tab:parameters}. The other planetary and orbital parameters are also consistent with the values presented in Table~\ref{tab:parameters}. For the GP hyper-parameters, we found $V_c=0.55 \pm 0.23$\,\ms, $V_r =70_{-21}^{+27}$\,\ms, $B_c  = 9.4_{-2.9}^{+3.4}$\,\ms, $B_r  = 64_{-25}^{+20}$\,\ms, $L_c = 5.4 \pm 2.2$\,\ms\, $P_{\rm GP} = 36.5 \pm 0.2$~d, $\lambda_{\rm e} = 4217_{-685}^{+624}$~d, and $\lambda_{\rm p} = 1086_{-394}^{+501}$. The relatively large values of the scale parameters in the GP, i.e. $\lambda_{\rm e}$ and $\lambda_{\rm p}$, indicate that the stellar activity behaves like a sinusoidal signal (with slight corrections).

\section{Discussion and conclusions}
\label{Sect.:DiscussionConclusions}

\begin{figure}
    \centering
    \includegraphics[width=0.47\textwidth]{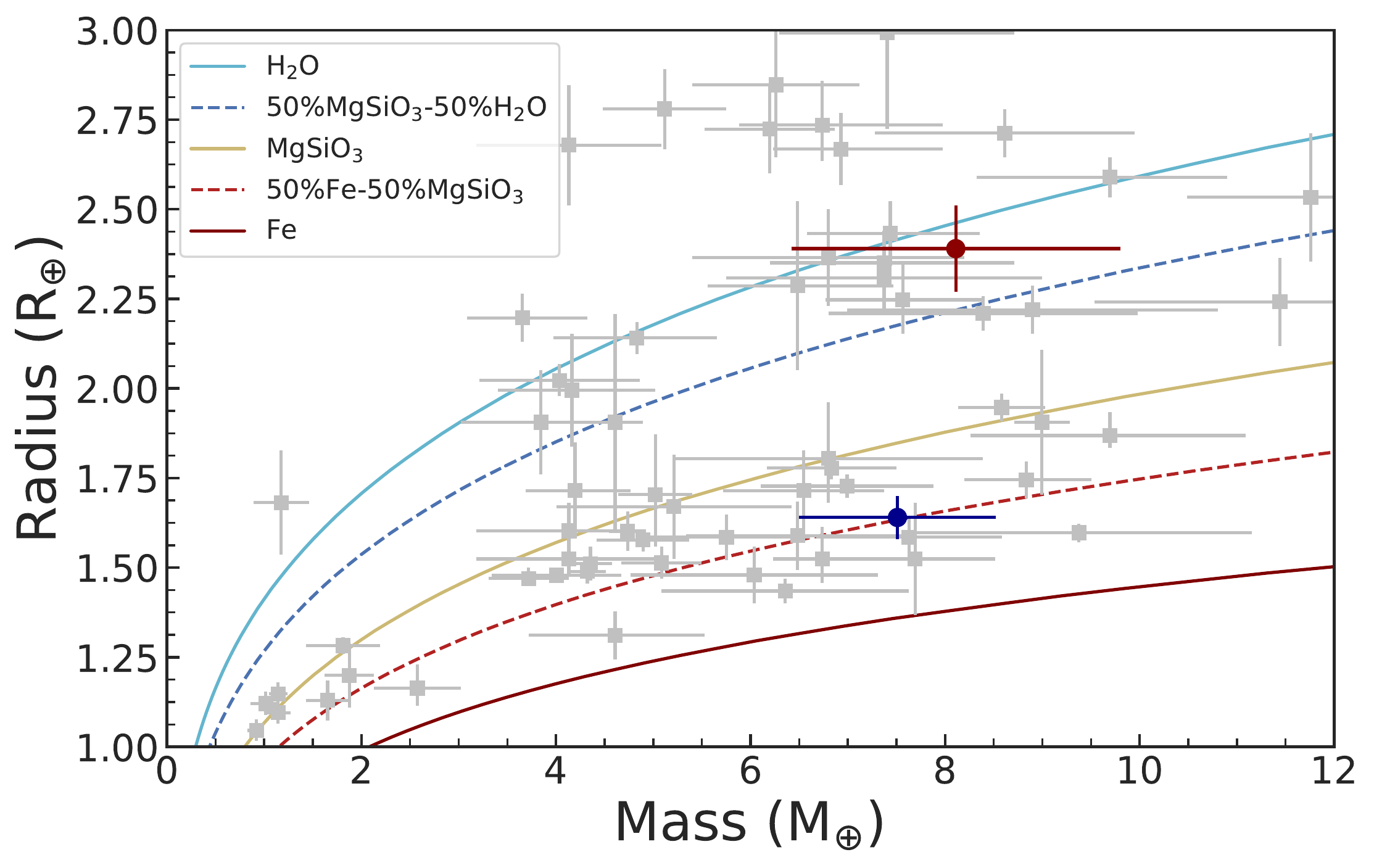}
    \caption{Mass--radius diagram for low-mass ($M_{\rm p}$\,$<$\,12~$M_\oplus$), small ($R_{\rm p}$\,$<$\,3\,$R_\oplus$) planets with mass-radius measurements better than 25\%  \citep[from \url{http://www.astro.keele.ac.uk/jkt/tepcat/};][]{Southworth2011}. Composition models from \citet{Zeng2016} are displayed with different lines and colors. The solid blue and red circles mark the position of \sname~b and \sname~c, respectively.
    }
    \label{Fig:MassRadiusDiagram}
\end{figure}

The innermost transiting planet \sname\,b ($P_\mathrm{orb,b} = 4.8$\,days) has a mass of $M_\mathrm{b}$=\mpb\ and a radius of $R_\mathrm{b}$=\rpb, yielding a mean density of $\rho_\mathrm{b}$=\denpb. Figure~\ref{Fig:MassRadiusDiagram} displays the position of \sname\,b on the mass-radius diagram compared to the sub-sample of small transiting planets ($R_\mathrm{p}\le3$~$R_\oplus$) whose masses and radii have been derived with a precision better than 25\%. Theoretical models from \citet{Zeng2016} are overplotted using different lines and colors. Given the precision of our mass determination ($\sim$14\%), we conclude that \sname\,b is a rocky terrestrial planet with a composition consisting of $\sim$50\,\% silicate and $\sim$50\% iron.

For \sname\,c ($P_\mathrm{orb,c}$=17.2\,days), we obtained a mass of $M_\mathrm{c}$=\mpc\ and a radius of $R_\mathrm{c}$=\rpc, yielding a mean density of $\rho_\mathrm{c}$=\denpc. Therefore, \sname\,b and c have similar masses, but the radius of \sname\,c is $\sim$1.5 times larger than the radius of \sname\,b. The lower bulk density of \sname\,c suggests that the planet is likely composed by a rocky core surrounded either by a considerable amount of water, or by a light, hydrogen-dominated envelope. In the case of a water-rich planet, the amount of water and high planetary equilibrium temperature would imply the presence of a steam atmosphere, which would be strongly hydrogen dominated in its upper layer, as a consequence of water dissociation and the low mass of hydrogen. It is therefore plausible to assume that \sname\,c hosts a hydrogen-dominated atmosphere, at least in its upper part.

As in other systems hosting two close-in sub-Neptune-mass planets \citep[e.g., HD\,3167][]{Gandolfi2017}, the radii of \sname\,b and c lie on opposite sides of the radius gap \citep{Fulton2017, VanEylen2018}, with the closer-in planet having a higher bulk density, similar to other close-in systems with measured planetary masses \citep[e.g., HD\,3167, K2-109, GJ\,9827;][]{Gandolfi2017, Guenther2017, Prieto-Arranz2018}. This gap is believed to be caused by atmospheric escape \citep{Owen2017,Jin2018}, which is stronger for closer-in planets. Within this picture, \sname\,b would probably have lost its primary, hydrogen-dominated atmosphere and now hosts a secondary atmosphere possibly resulting from out-gassing of a solidifying magma ocean, while \sname\,c is likely to still partly retain the primordial hydrogen-dominated envelope. This is consistent with \citet{VanEylen2018}, who measured the location and slope of the radius gap as a function of orbital period and matched it to models suggesting a homogeneous terrestrial core composition.

To first order, the radii of \sname\,b and c depend on the present-day properties of their atmospheres, which are intimately related to the amount of high-energy (X-ray and extreme ultraviolet; $\lambda < 91.2$\,nm) stellar radiation received since the dispersal of the protoplanetary nebula, and thus also to the stellar rotation history. The evolution of the stellar rotation rate does not follow a unique path because stars of the same mass and metallicity can have significantly different rotation rates up to about 1\,Gyr \citep[e.g.,][]{Mamajek2008,Johnstone2015,Tu2015}. For older stars, it is therefore not possible to infer their past high-energy emission from their measured stellar properties. Starting from the assumption that \sname\,c hosted a hydrogen-dominated atmosphere with solar metallicity throughout its entire evolution, we derived the history of the stellar rotation and high-energy emission by modeling the atmospheric evolution of \sname\,c. To this end, we employed the atmospheric evolution algorithm described by \citet{Kubyshkina2018} and further developed by Kubyshkina et al. (2019, ApJ, in press), which is based on a Bayesian approach, fitting the currently observed planetary radius and combining the planetary evolution model with the MCMC open-source algorithm of \citet{Cubillos2017}. The planetary atmospheric evolution model, system parameters (i.e., planetary mass, planetary radius, orbital separation, current stellar rotation period, stellar age, stellar mass; Table~\ref{tab:parameters}) were then used to compute the posterior distribution for the stellar rotation rate at any given age via MCMC. We assumed Gaussian priors determined by the measured system parameters and their uncertainties.

\begin{figure}
    \centering
    \includegraphics[width=0.49\textwidth]{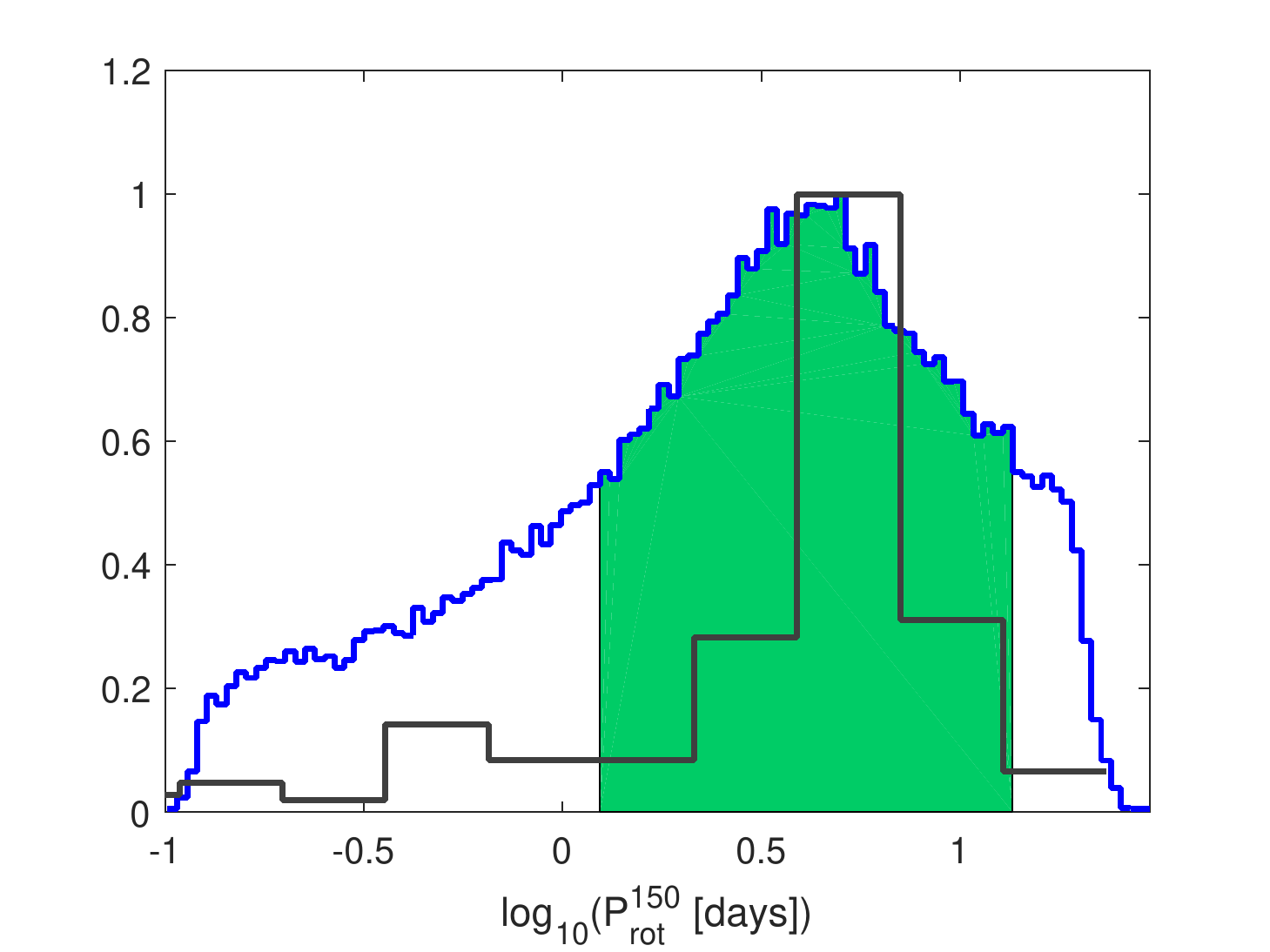}
    \caption{MCMC posterior distributions for the stellar rotation period at an age of 150\,Myr obtained from the modeling of \sname\,c. The shaded areas correspond to the 68\% region of the credible interval of the posterior distribution. The black histogram shows the distribution of stellar rotation periods measured for open cluster stars with an age of 150\,Myr \citep[from][]{Johnstone2015}.
    }
    \label{Fig:starevol}
\end{figure}

Figure~\ref{Fig:starevol} shows the obtained posterior distribution for the rotation period \sname\ at an age of 150\,Myr in comparison with the distribution derived from measurements of open cluster stars of the same age \citep{Johnstone2015}. Our results indicate that \sname, when it was young, was likely to be a moderate rotator, with a high-energy emission at 150\,Myr ranging between 3.7 and 127 times the current solar emission, further excluding that the star was a very fast/slow rotator. We further employed the result shown in Fig.~\ref{Fig:starevol} to estimate the past atmospheric evolution of a possible hydrogen-dominated atmosphere of \sname\,b. Accounting for all uncertainties on the system parameters and on the derived history of the stellar rotation period, we obtained that \sname\,b has completely lost its primary atmosphere, assuming it held one, within 300\,Myr, in agreement with the currently observed mean density.

The position of \sname\,c in the mass-radius diagram (Fig.~\ref{Fig:MassRadiusDiagram}) indicates that the planet may be hosting a massive hydrogen-dominated envelope or a smaller secondary atmosphere. As primary atmospheres are easily subject to escape, knowing the current composition of the envelope of \sname\,c would provide a strong constraint on atmospheric evolution models. In this respect, this planet is similar to $\pi$\,Men\,c \citep{Gandolfi2018,Huang2018b}; furthermore, as for $\pi$\,Men, the close distance to the system and brightness of the host star would enable high-quality transmission spectroscopy spanning from far-ultraviolet to infrared wavelengths. Of particular interest would be probes of an extended, escaping atmosphere. Spectral lines sensitive to various levels of extended atmospheres include H\,{\sc i}, C\,{\sc ii}, and O\,{\sc i} resonance lines in the ultraviolet, H$_\alpha$ in the optical, and He\,{\sc I} in the near-infrared. This suite of lines would provide a comprehensive picture of the upper atmosphere of the planet, thus constraining atmospheric escape and evolution models.

\begin{table*}[!t]
 \scriptsize
\begin{center}
  \caption{\sname\ System Parameters. \label{tab:parameters}}  
  \begin{tabular}{lcc}
  \hline\hline
  \noalign{\smallskip}
  Parameter & Prior$^{(\mathrm{a})}$ & Derived Value \\
  \noalign{\smallskip}
  \hline
    \noalign{\smallskip}
    \multicolumn{3}{l}{\emph{\bf{Stellar Parameters}}} \\
    \noalign{\smallskip}
    Star mass $M_{\star}$ ($M_\odot$)             &  $\cdots$ & \smass[]    \\
    Star radius $R_{\star}$ ($R_\odot$)           &  $\cdots$ & \sradius[]  \\
    Effective temperature $\mathrm{T_{eff}}$ (K)  & $\cdots$  & \stemp[]    \\
    Surface gravity$^{(\mathrm{b})}$ \logg\ (cgs) & $\cdots$ & 4.53\,$\pm$\,0.02 \\
    Surface gravity$^{(\mathrm{c})}$ \logg\ (cgs) & $\cdots$ & 4.40\,$\pm$\,0.10 \\
    Iron abundance [Fe/H] (dex)                   & $\cdots$ & 0.15\,$\pm$\,0.10 \\
    Sodium abundance [Na/H] (dex)                 & $\cdots$ & 0.27\,$\pm$\,0.09 \\
    Calcium abundance [Ca/H] (dex)                & $\cdots$ & 0.16\,$\pm$\,0.05 \\
    Projected rotational velocity \vsini\ (\kms)  & $\cdots$ & 1.0\,$\pm$\,1.0   \\
    Age (Gyr)                                     & $\cdots$ & 5.1\,$\pm$\,0.8   \\
    Interstellar extinction $A_\mathrm{v}$        & $\cdots$ & 0.02\,$\pm$\,0.02 \\
    \noalign{\smallskip}
    \hline
    \noalign{\smallskip}
    \multicolumn{3}{l}{\emph{\bf{Model Parameters of \sname~b}}} \\
    \noalign{\smallskip}
    Orbital period $P_{\mathrm{orb,\,b}}$ (days) &  $\mathcal{U}[4.7552 , 4.7572]$ & \Pb[] \\
    Transit epoch $T_\mathrm{0,\,b}$ (BJD$_\mathrm{TDB}-$2\,450\,000) & $\mathcal{U}[8411.4526 , 8411.4706]$ & \Tzerob[]  \\ 
    Scaled semi-major axis $a_\mathrm{b}/R_{\star}$ &  $\mathcal{N}[13.11,0.17]$ & \arb[] \\
    Planet-to-star radius ratio $R_\mathrm{b}/R_{\star}$ & $\mathcal{U}[0,0.1]$ & \rrb[]  \\
    Impact parameter $b_\mathrm{b}$  & $\mathcal{U}[0,1]$  & \bb[] \\
    $\sqrt{e_\mathrm{b}} \sin \omega_\mathrm{\star,\,b}$ &  $\mathcal{U}[-1,1]$ & \esinb[] \\
    $\sqrt{e_\mathrm{b}} \cos \omega_\mathrm{\star,\,b}$ &  $\mathcal{U}[-1,1]$ & \ecosb[] \\
    RV semi-amplitude variation $K_\mathrm{b}$ (\ms) & $\mathcal{U}[0,10]$ & \kb[] \\
    \noalign{\smallskip}
    \hline
    \noalign{\smallskip}
    \multicolumn{3}{l}{\emph{\bf{Model Parameters of \sname~c}}} \\
    \noalign{\smallskip}
    Orbital period $P_{\mathrm{orb,\,c}}$ (days) &  $\mathcal{U}[ 17.1676 , 17.1876 ]$ & \Pc[] \\
    Transit epoch $T_\mathrm{0,\,c}$ (BJD$_\mathrm{TDB}-$2\,450\,000) & $\mathcal{U}[8414.5416 , 8414.5616]$ & \Tzeroc[]  \\     
    Scaled semi-major axis $a_\mathrm{c}/R_{\star}$ &  $\mathcal{N}[31.68,0.70]$ & \arc[] \\
    Planet-to-star radius ratio $R_\mathrm{c}/R_{\star}$ & $\mathcal{U}[0,0.1]$ & \rrc[]  \\
    Impact parameter $b_\mathrm{c}$  & $\mathcal{U}[0,1]$  & \bc[] \\
    $\sqrt{e_\mathrm{c}} \sin \omega_\mathrm{\star,\,c}$ &  $\mathcal{U}[-1,1]$ & \esinc[] \\
    $\sqrt{e_\mathrm{c}} \cos \omega_\mathrm{\star,\,c}$ &  $\mathcal{U}[-1,1]$ & \ecosc[] \\
    RV semi-amplitude variation $K_\mathrm{c}$ (\ms) & $\mathcal{U}[0,10]$ & \kc[] \\
    \noalign{\smallskip}
    \hline
    \noalign{\smallskip}
    \multicolumn{3}{l}{\emph{\bf{Additional Model Parameters}}} \\
    \noalign{\smallskip}
    Parameterized limb-darkening coefficient $q_1$  & $\mathcal{N}[0.43,0.1]$ & \qone \\
    Parameterized limb-darkening coefficient $q_2$  & $\mathcal{N}[0.19,0.1]$ & \qtwo \\
    Systemic velocity $\gamma_{\mathrm{HS1}}$  (km s$^{-1}$) & $\mathcal{U}[$-$4.0 , $-$3.6]$ & \HARPSone[]  \\
    Systemic velocity $\gamma_{\mathrm{HS2}}$  (km s$^{-1}$) & $\mathcal{U}[$-$4.0 , $-$3.6]$ & \HARPStwo[] \\
    RV jitter term $\sigma_{\mathrm{HS1}}$  (\ms) & $\mathcal{U}[0,100]$ & \jHARPSone[]  \\
    RV jitter term $\sigma_{\mathrm{HS2}}$  (\ms) & $\mathcal{U}[0,100]$ & \jHARPStwo[]  \\
    Stellar rotation period ($P_\mathrm{rot}$) days & $\mathcal{U}[36.4,36.6]$ & \Pd[] \\
    Linear RV term ${\rm m\,s^{-1}\,d^{-1}}$ & $\mathcal{U}[-0.1,0.1]$& \ltrend[] \\
    Quadratic RV term  ${\rm m\,s^{-1}\,d^{-1}}$ & $\mathcal{U}[-0.1,0.1]$ & \qtrend[] \\
    \noalign{\smallskip}
    \hline 
    \noalign{\smallskip}
    \multicolumn{3}{l}{\emph{\bf{Derived Parameters of \sname\,b}}} \\
    \noalign{\smallskip}

    Planet mass $M_\mathrm{b}$ ($M_\oplus$) & $\cdots$ & \mpb[]  \\
    Planet radius $R_\mathrm{b}$ ($R_{\oplus}$) & $\cdots$ & \rpb[] \\
    Planet mean density $\rho_\mathrm{b}$ ($\mathrm{g\,cm^{-3}}$) & $\cdots$ & \denpb[] \\
    Semi-major axis of the planetary orbit $a_\mathrm{b}$ (au) & $\cdots$ & \ab[]  \\
    Orbit eccentricity $e_\mathrm{b}$ & $\cdots$ & \eb \\
    Argument of periastron of stellar orbit $\omega_\mathrm{\star,\,b}$ (deg) & $\cdots$ & \wb[] \\
    Orbit inclination $i_\mathrm{b}$ (deg) & $\cdots$ & \ib[] \\
    Equilibrium temperature$^{(\mathrm{d})}$  $T_\mathrm{eq,\,b}$ (K)  & $\cdots$ &  \Teqb[] \\
    Transit duration $\tau_\mathrm{14,\,b}$ (hr) & $\cdots$ & \ttotb[] \\    \noalign{\smallskip}
    \hline
    \noalign{\smallskip}
    \multicolumn{3}{l}{\emph{\bf{Derived Parameters of \sname\,c}}} \\
    \noalign{\smallskip}

    Planet mass $M_\mathrm{c}$ ($M_{\oplus}$) & $\cdots$ & \mpc[]  \\
    Planet radius $R_\mathrm{c}$ ($R_{\oplus}$) & $\cdots$ & \rpc[] \\
    Planet mean density $\rho_\mathrm{c}$ ($\mathrm{g\,cm^{-3}}$) & $\cdots$ & \denpc[] \\
    Semi-major axis of the planetary orbit $a_\mathrm{c}$ (au) & $\cdots$ & \ac[]  \\
    Orbit eccentricity $e_\mathrm{c}$ & $\cdots$ & \ec[] \\
    Argument of periastron of stellar orbit $\omega_\mathrm{\star,\,c}$ (deg) & $\cdots$ & \wc[] \\
    Orbit inclination $i_\mathrm{c}$ (deg) & $\cdots$ & \ic[] \\
    Equilibrium temperature$^{(\mathrm{d})}$  $T_\mathrm{eq,\,c}$ (K)  & $\cdots$ &  \Teqc[] \\
    Transit duration $\tau_\mathrm{14,\,c}$ (hr) & $\cdots$ & \ttotc[] \\
    \noalign{\smallskip}
  \hline
  \end{tabular}
\end{center}
\tablecomments{(a) $\mathcal{U}[a,b]$ refers to uniform priors between $a$ and $b$; $\mathcal{N}[a,b]$ to Gaussian priors with mean $a$ and standard deviation $b$; (b) from spectroscopy and evolutionary tracks; (c) from spectroscopy; (d) assuming zero albedo and uniform~redistribution~of~heat.}
\end{table*}

\begin{table*}
\footnotesize
\begin{center}
\caption{\label{Table:HARPS1} HARPS RV measurements of \sname\ acquired with the old fiber bundle.}
\begin{tabular}{lcccccr}
\hline\hline
\noalign{\smallskip}
$\rm BJD_{TDB}^a$ & RV  & $\pm \sigma$ &    BIS  & FWHM   & T$_\mathrm{exp}$ & S/N $^b$\\
-2450000          & (\kms)  & (\kms)       &  (\kms) & (\kms)  &      (s)         &   \\
\hline
\noalign{\smallskip}
\noalign{\smallskip}
 2988.663700 & -3.8208 & 0.0010 &  0.0013 & 6.1330 &  900 &  70.8 \\
 3270.822311 & -3.8260 & 0.0008 &  0.0015 & 6.1353 &  900 &  93.7 \\
 3785.541537 & -3.8234 & 0.0007 &  0.0010 & 6.1471 &  900 & 101.7 \\
 4422.673842 & -3.8155 & 0.0006 &  0.0059 & 6.1571 &  900 & 108.1 \\
 4424.646720 & -3.8111 & 0.0007 &  0.0037 & 6.1531 &  900 &  97.7 \\
 4427.703292 & -3.8185 & 0.0006 &  0.0049 & 6.1547 &  900 & 101.4 \\
 4428.644416 & -3.8170 & 0.0005 &  0.0034 & 6.1481 &  900 & 125.8 \\
 4484.550086 & -3.8203 & 0.0006 &  0.0063 & 6.1631 &  900 & 102.6 \\
 4730.822010 & -3.8128 & 0.0006 &  0.0100 & 6.1779 &  900 & 108.1 \\
 4731.764597 & -3.8127 & 0.0007 &  0.0133 & 6.1767 &  900 &  98.8 \\
 4734.786220 & -3.8132 & 0.0010 &  0.0150 & 6.1661 &  900 &  68.4 \\
 4737.774983 & -3.8115 & 0.0011 &  0.0053 & 6.1684 &  900 &  62.2 \\
 4739.782074 & -3.8140 & 0.0008 &  0.0015 & 6.1633 & 1200 &  79.4 \\
 4801.645505 & -3.8101 & 0.0006 &  0.0061 & 6.1753 &  900 & 112.5 \\
 4802.681221 & -3.8117 & 0.0007 &  0.0065 & 6.1745 &  900 & 100.7 \\
 4803.585301 & -3.8107 & 0.0006 &  0.0078 & 6.1681 &  900 & 117.0 \\
 4804.621351 & -3.8088 & 0.0008 &  0.0102 & 6.1717 &  900 &  85.6 \\
 4806.641716 & -3.8139 & 0.0006 &  0.0078 & 6.1734 &  900 & 103.1 \\
 4847.567925 & -3.8116 & 0.0006 &  0.0062 & 6.1635 &  900 & 113.6 \\
 5038.928157 & -3.8169 & 0.0006 &  0.0088 & 6.1583 &  900 & 113.4 \\
 5039.878459 & -3.8165 & 0.0009 &  0.0028 & 6.1577 &  900 &  78.1 \\
 5040.884957 & -3.8191 & 0.0016 & -0.0011 & 6.1644 &  900 &  47.0 \\
 5042.901725 & -3.8105 & 0.0005 &  0.0036 & 6.1476 &  900 & 120.8 \\
 5067.879903 & -3.8164 & 0.0007 &  0.0068 & 6.1558 &  900 &  93.6 \\
 5068.916098 & -3.8185 & 0.0008 &  0.0056 & 6.1655 &  800 &  85.2 \\
 5070.833766 & -3.8159 & 0.0008 &  0.0071 & 6.1630 &  900 &  84.5 \\
 5097.828049 & -3.8123 & 0.0008 &  0.0012 & 6.1615 &  900 &  79.2 \\
 5100.771149 & -3.8063 & 0.0007 &  0.0049 & 6.1668 &  900 &  91.2 \\
 5106.752698 & -3.8165 & 0.0009 &  0.0080 & 6.1621 &  900 &  76.6 \\
 5108.758136 & -3.8147 & 0.0009 &  0.0105 & 6.1507 &  900 &  73.4 \\
 5110.725697 & -3.8138 & 0.0007 &  0.0023 & 6.1464 &  900 &  89.9 \\
 5113.727962 & -3.8110 & 0.0006 &  0.0041 & 6.1508 &  900 & 116.6 \\
 5116.732322 & -3.8188 & 0.0008 &  0.0028 & 6.1553 &  900 &  85.7 \\
 5124.719074 & -3.8125 & 0.0005 &  0.0052 & 6.1507 &  900 & 126.2 \\
 5134.807289 & -3.8104 & 0.0007 &  0.0066 & 6.1632 &  900 &  94.7 \\
 5137.624046 & -3.8095 & 0.0006 &  0.0115 & 6.1656 &  900 & 102.7 \\
 5141.642265 & -3.8120 & 0.0006 &  0.0095 & 6.1629 &  900 & 111.4 \\
 5164.557710 & -3.8122 & 0.0006 &  0.0038 & 6.1587 &  900 & 104.4 \\
 5166.557368 & -3.8099 & 0.0006 &  0.0008 & 6.1533 &  900 & 104.2 \\
 5169.552068 & -3.8161 & 0.0005 &  0.0008 & 6.1593 &  900 & 120.8 \\
 5227.530636 & -3.8150 & 0.0007 &  0.0091 & 6.1510 &  900 &  98.7 \\
 5230.529883 & -3.8161 & 0.0007 & -0.0025 & 6.1456 &  900 &  94.4 \\
 5245.518763 & -3.8120 & 0.0007 &  0.0087 & 6.1559 &  900 &  97.8 \\
 $^*$5246.519846$^*$ & -3.8169 & 0.0360 &  0.1136 & 6.3367 &    5 &   3.8 \\
 5246.526257 & -3.8094 & 0.0007 &  0.0047 & 6.1576 &  900 &  94.9 \\
 6620.642369 & -3.8098 & 0.0007 & -0.0005 & 6.1539 & 1200 &  92.8 \\
 6623.580646 & -3.8089 & 0.0010 &  0.0033 & 6.1612 &  900 &  69.8 \\
 6625.634172 & -3.8073 & 0.0008 &  0.0024 & 6.1609 &  900 &  92.2 \\
 6628.589169 & -3.8119 & 0.0013 &  0.0010 & 6.1674 &  900 &  58.4 \\
 6631.568567 & -3.8060 & 0.0008 &  0.0040 & 6.1663 &  900 &  90.9 \\
 7036.603445 & -3.8091 & 0.0009 &  0.0001 & 6.1772 &  900 &  83.9 \\
 7037.560130 & -3.8147 & 0.0008 &  0.0041 & 6.1679 &  900 &  88.9 \\
 \noalign{\smallskip}
\hline
\end{tabular}
\\
\end{center}
\tablecomments{(a) Barycentric Julian dates are given in barycentric dynamical time;
(b) S/N per pixel at 550 nm; (*) outlier not included in the analysis.\\}
\end{table*}

\begin{table*}
\footnotesize
\begin{center}
\caption{\label{Table:HARPS2} HARPS RV measurements of \sname\ acquired with the new fiber bundle.}
\begin{tabular}{lcccccr}
\hline\hline
\noalign{\smallskip}
$\rm BJD_{TDB}^a$ & RV  & $\pm \sigma$ &    BIS  & FWHM   & T$_\mathrm{exp}$ & S/N $^b$\\
-2450000          & (\kms)  & (\kms)       &  (\kms) & (\kms)  &      (s)         &   \\
\hline
\noalign{\smallskip}
\noalign{\smallskip}
 7291.826359 & -3.7848 & 0.0008 &  0.0273 & 6.2102 &  900 &  88.1 \\
 7292.799439 & -3.7889 & 0.0007 &  0.0246 & 6.2141 &  900 &  95.6 \\
 7299.843584 & -3.7988 & 0.0007 &  0.0229 & 6.1968 &  900 & 106.2 \\
 7303.879389 & -3.7973 & 0.0013 &  0.0184 & 6.1985 &  900 &  60.3 \\
 7353.698805 & -3.7959 & 0.0008 &  0.0209 & 6.2081 &  900 &  97.6 \\
 7357.682828 & -3.7898 & 0.0006 &  0.0239 & 6.2073 &  900 & 127.1 \\
 7373.685080 & -3.7889 & 0.0007 &  0.0207 & 6.2064 &  900 &  99.5 \\
 7395.644348 & -3.7931 & 0.0011 &  0.0253 & 6.2093 &  900 &  68.5 \\
 7399.617257 & -3.7938 & 0.0008 &  0.0213 & 6.1981 &  900 &  90.1 \\
 7418.584180 & -3.7951 & 0.0007 &  0.0267 & 6.1927 &  900 & 108.5 \\
 7422.589672 & -3.7969 & 0.0008 &  0.0211 & 6.1964 &  900 & 103.2 \\
 7427.538475 & -3.7932 & 0.0008 &  0.0197 & 6.1990 &  900 &  94.5 \\
 7429.539077 & -3.7883 & 0.0006 &  0.0222 & 6.2054 &  900 & 129.3 \\
 7584.927622 & -3.7969 & 0.0007 &  0.0218 & 6.2010 &  900 &  97.2 \\
 7613.935104 & -3.8009 & 0.0007 &  0.0214 & 6.1960 &  900 & 104.8 \\
 $^*$7641.794439$^*$ & -1.6179 & 0.0009 &  0.0220 & 6.1919 &  900 &  81.4 \\
 7642.837840 & -3.8015 & 0.0009 &  0.0196 & 6.1934 &  900 &  80.2 \\
 7643.808690 & -3.8000 & 0.0008 &  0.0189 & 6.1928 &  900 &  89.1 \\
 7644.862657 & -3.7982 & 0.0009 &  0.0202 & 6.1907 &  900 &  79.1 \\
 7647.923051 & -3.7972 & 0.0007 &  0.0181 & 6.1888 &  900 & 110.7 \\
 7649.726575 & -3.7895 & 0.0010 &  0.0205 & 6.1981 &  900 &  72.2 \\
 7650.752860 & -3.7912 & 0.0006 &  0.0207 & 6.2017 &  900 & 115.2 \\
 7652.744706 & -3.7889 & 0.0008 &  0.0222 & 6.1982 &  900 &  91.5 \\
 7656.751475 & -3.7956 & 0.0008 &  0.0254 & 6.2022 &  900 &  87.1 \\
 7658.854305 & -3.7910 & 0.0005 &  0.0196 & 6.1970 &  900 & 142.1 \\
 7660.797461 & -3.7952 & 0.0008 &  0.0195 & 6.1985 &  900 &  85.9 \\
 7661.831637 & -3.7959 & 0.0007 &  0.0116 & 6.2990 &  900 & 107.0 \\
 7971.834240 & -3.7981 & 0.0009 &  0.0219 & 6.1895 &  900 &  79.6 \\
 7993.916286 & -3.8000 & 0.0009 &  0.0246 & 6.2071 &  900 &  84.7 \\
 7994.887034 & -3.7979 & 0.0011 &  0.0250 & 6.2032 &  900 &  70.0 \\
 7996.829372 & -3.7884 & 0.0012 &  0.0185 & 6.2004 &  900 &  64.2 \\
 7996.923718 & -3.7902 & 0.0009 &  0.0238 & 6.1989 & 1500 &  89.6 \\
 7998.866747 & -3.7986 & 0.0010 &  0.0216 & 6.2004 &  900 &  72.2 \\
 8001.872927 & -3.7930 & 0.0008 &  0.0188 & 6.1873 &  900 &  99.5 \\
 8002.895001 & -3.7975 & 0.0009 &  0.0216 & 6.1880 &  900 &  82.8 \\
\noalign{\smallskip}
\hline
\end{tabular}
\\
\end{center}
\tablecomments{(a) Barycentric Julian dates are given in barycentric dynamical time;
(b) S/N per pixel at 550 nm; (*) outlier not included in the analysis.\\}
\end{table*}

\acknowledgements

We are very grateful to the referee for a careful reading of this Letter and valuable suggestions and comments. We acknowledge the use of public \tess\ Alert data from pipelines at the \tess\ Science Office and at the \tess\ Science Processing Operations Center. This Letter includes data collected by the \tess\ mission, which are publicly available from the Mikulski Archive for Space Telescopes (MAST). Resources supporting this work were provided by the NASA High-End Computing (HEC) Program through the NASA Advanced Supercomputing (NAS) Division at Ames Research Center for the production of the SPOC data products. Funding for the \tess\ mission is provided by NASA's Science Mission directorate. Based on observations collected at the European Organization for Astronomical Research in the Southern Hemisphere under ESO programs 072.C-0488, 183.C-0972, 192.C-0852, 196.C-1006, and 198.C-0836. This research has made use of the services of the ESO Science Archive Facility. L.F. and D.K. acknowledge the Austrian Forschungsförderungsgesellschaft FFG project ``TAPAS4CHEOPS'' P853993. J.H.L. acknowledges the support of the Japan Society for the Promotion of Science (JSPS) Research Fellowship for Young Scientists. J.K., S.G., M.P., S.C., K.W.F.L., H.R., A.H., and M.E. acknowledge the support by DFG Grants PA525/18-1, PA525/19-1, PA-525/20-1, HA 3279/12-1 and RA 714/14-1 within the DFG Priority Program SPP1992: ``Exploring the Diversity of Exoplanets.'' H.J.D. and D.N. acknowledge support by grants ESP2015-65712-C5-4-R, ESP2016-80435-C2-2-R, and ESP2017-87676-C5-4-R of the Spanish Secretary of State for R\&D\&i (MINECO). S.C. thanks the Hungarian National Research, Development and Innovation Office for the NKFI-KH-130372 grant. SM acknowledges support from the Ramon y Cajal fellowship number RYC-2015-17697. This work is partly supported by JSPS KAKENHI grant Nos. JP18H01265 and 18H05439, and JST PRESTO grant No. JPMJPR1775. I.R. acknowledges support from the Spanish Ministry for Science, Innovation and Universities (MCIU) and the Fondo Europeo de Desarrollo Regional (FEDER) through grant ESP2016-80435-C2-1-R, as well as the support of the Generalitat de Catalunya/CERCA programme. M.F. and C.M.P. gratefully acknowledge the support of the Swedish National Space Agency. 17-01752J. M.S. acknowledges the Postdoc@MUNI project CZ.02.2.69/0.0/0.0/16-027/0008360. 

\facilities{\tess, HARPS}
\software{{\tt VARLET} \citep{Grziwa2016}, {\tt PHALET} \citep{Grziwa2016}, {\tt SME} \citep{{vp96,vf05,Piskunov2017}},
{\tt pyaneti} \citep{Barragan2019}}

\bibliographystyle{aasjournal}
\bibliography{bibliography}

\end{document}